\RequirePackage{fix-cm}
\documentclass[twocolumn]{svjour3}          
\smartqed  
\usepackage{graphicx}
\usepackage[lined,ruled]{algorithm2e}
\usepackage{algorithmic}
\usepackage{subfig}
\usepackage{amsmath} 
\usepackage{multirow}
\usepackage{url}
\usepackage{color}
\usepackage{enumitem}

\usepackage{amssymb}
\usepackage{amsthm}
\newcommand{\IGNORE}[1]{}
\newcommand{\vv}[1]{\ensuremath{\mathbf{#1}}}
\newcommand{\vel}[1]{\ensuremath{\mathbf{v}_{#1}}}
\newcommand{\vnew}[1]{\ensuremath{\mathbf{v}^\mathrm{new}_{#1}}}
\newcommand{\vpref}[1]{\ensuremath{\mathbf{v}^\mathrm{pref}_{#1}}}
\newcommand{\vgoal}[1]{\ensuremath{\mathbf{v}^\mathrm{goal}_{#1}}}
\newcommand{\umax}[1]{\ensuremath{\upsilon^{\max}_{#1}}\xspace}

\newcommand{\radius}[1]{\ensuremath{r_{#1}}\xspace}

\newcommand{\agent}[1]{\ensuremath{A_{#1}}\xspace}
\newcommand{\pos}[1]{\ensuremath{\vv{p}_{#1}}\xspace}
\newcommand{\post}[1]{\ensuremath{\vv{p}_{#1}^\mathrm{\textit{t}}}\xspace}
\newcommand{\postt}[1]{\ensuremath{\vv{p}_{#1}^\mathrm{\textit{t-1}}}\xspace}
\newcommand{\goal}[1]{\ensuremath{\vv{g}_{#1}}\xspace}

\DeclareMathOperator*{\argmin}{arg\,min}

\journalname{Autonomous Robots}
\begin{document}

\title{ALAN: Adaptive Learning for Multi-Agent Navigation
}


\author{Julio Godoy         \and
        Tiannan Chen \and
        Stephen J. Guy \and
        Ioannis Karamouzas \and
        Maria Gini 
}


\institute{Julio Godoy\at
              Department of Computer Science, Universidad de Concepcion \\
              Edmundo Larenas 219, Concepcion, Chile\\
              \email{juliogodoy@gmail.com}           
           \and
           Tiannan Chen, Stephen J. Guy and Maria Gini \at
           Department of Computer Science and Engineering, University of Minnesota \\
             200 Union Street SE, Minneapolis, MN 55455, USA
          \and    
           Ioannis Karamouzas \at
           School of Computing, Clemson University\\
           100 McAdams Hall, Clemson, South Carolina, SC 29634, USA
}

\date{Received: date / Accepted: date}

\maketitle

\begin{abstract}
In multi-agent navigation, agents need to move towards their goal locations while avoiding collisions with other agents and static obstacles, often without communication with each other. Existing methods compute motions that are optimal locally but do not account for the aggregated motions of all agents, producing inefficient global behavior especially when agents move in a crowded space. 
In this work, we develop methods to allow agents to dynamically adapt their behavior to their local conditions. We accomplish this by formulating the multi-agent navigation problem as an action-selection problem, and propose an approach, ALAN, that allows agents to compute time-efficient and collision-free motions. ALAN is highly scalable because each agent makes its own decisions on how to move using a set of velocities optimized for a variety of navigation tasks. Experimental results show that the agents using ALAN, in general, reach their destinations faster than using ORCA, a state-of-the-art collision avoidance framework, the Social Forces model for pedestrian navigation, and a Predictive collision avoidance model.

\keywords{Multi-agent navigation \and Online learning \and Action selection \and Multi-agent coordination}
\end{abstract}

\section{Introduction}

Real-time goal-directed navigation of multiple agents is required in many domains, such as swarm robotics, pedestrian
navigation, planning for evacuation, and traffic engineering.
Conflicting constraints and the need to operate in real time make this problem challenging.
Agents need to move towards their goals in a timely manner, but also need to avoid collisions with each other and the environment. In addition, due to the number of agents and the
real-time constraints, each agent needs to compute its own motion without any communication with the other agents. 

While decentralization is essential for scalability and robustness, achieving globally efficient motions is critical, especially in applications such as search and rescue, aerial surveillance, and evacuation planning, where time is critical. 
Over the past twenty years, many decentralized techniques for real-time multi-agent navigation have been proposed, with approaches such as Optimal Reciprocal Collision Avoidance (ORCA)
~\cite{Berg09} being able to provide guarantees about collision-free motion for the agents. Although such techniques generate
locally efficient motions for each agent, the overall flow and global behavior of the agents can be far from efficient; agents plan only for themselves and do not consider how their motions affect the other agents. This leads to inefficient motions, congestion, and even deadlocks.

In this paper, we are interested in situations where agents have to minimize their overall travel time. 
We assume each agent has a preferred velocity indicating its desired direction of motion (typically oriented towards its goal) and speed. 
An agent runs a continuous cycle of sensing and acting. In each cycle, it has to choose a new velocity that avoids obstacles but is as close as possible to its preferred velocity. 
We show that by intelligently selecting preferred velocities that account for the global state of the multi-agent system, the time efficiency of the 
entire crowd can be significantly improved. 

In our setting, agents learn how to select their velocities in an online fashion without communicating with each other. 
To do so, we adapt a multi-armed bandit formulation to the velocity selection problem and present ALAN (Adaptive Learning Approach for Multi-Agent Navigation). With ALAN, agents choose intelligently from a set of actions, one at each timestep, based on both their goal and on how their motion will affect other agents. We show how critical the set of available actions is to performance, and we present a Markov Chain Monte Carlo learning method to learn an optimized action space for navigation in a variety of environments. Together with a scheme that guarantees collision-free motions, these features allow ALAN agents to minimize their overall travel time. \footnote{Videos highlighting our work can be found in \mbox{\fontsize{7.5}{7.5}\url{ http://motion.cs.umn.edu/r/ActionSelection}}}

\textbf{Main Results.} This paper presents four main contributions.
First, we formulate the 
multi-agent navigation problem in a multi-armed bandit setting. This enables each agent to decide its motions independently of the other agents. The other agents affect indirectly how an agent moves, because they affect the reward the agent receives. The independence of the choices made by each agent makes
the approach highly scalable.
Second, we propose an online action selection method inspired by the Softmax action selection technique \cite{SB98}, which
achieves the exploration exploitation tradeoff.
Third, we propose a Markov Chain Monte Carlo method to learn offline an optimized action set for specific navigation environments, as well as an action set optimized for multiple navigation scenarios. 
Last, we show experimentally that our approach leads to more time efficient motions in a variety of scenarios, reducing the travel time of all agents as compared to ORCA, the Social Forces model for simulating pedestrian dynamics \cite{helbing1995social} and the Pedestrian model for collision avoidance \cite{KHB+09}.

This work is an extended version of \cite{godoy2015adaptive}, which introduced a multi-armed bandit formulation for multi-agent navigation problems. 
Compared to \cite{godoy2015adaptive}, here we reduce ALAN's dependency on parameters, present 
an offline approach to learn an optimized action set, and perform an extended experimental analysis of ALAN.

The rest of the paper is organized as follows. In Section \ref{sec:related},
we review relevant related work. 
In Section \ref{sec:bgd}, we provide background on collision avoidance methods, especially on ORCA which is used in ALAN. In Section \ref{sec:man}, we present our problem formulation for multi-agent navigation. ALAN and its components are described in
Section \ref{sec:mainalan}, while our experimental setup is described in Section \ref{sec:quant}, where we also present our performance metric, the scenarios we use to
evaluate our approach, and experimental results. Section \ref{sec:mcmc} presents our Markov Chain Monte Carlo method for learning action spaces for different navigation environments. We perform a thorough experimental analysis of the performance of ALAN in Section \ref{sec:sens}, where we also discuss its applicability in multi-robot systems. 
Finally, we conclude and present future research plans in Section \ref{sec:conc}.

\section{Related Work}
\label{sec:related}

Extensive research in the areas of multi-agent navigation and learning has been conducted over the last decade. In this section, we present an overview of prior work most closely related to our approach. For a more comprehensive discussion on multi-agent navigation and learning we refer the reader to the surveys of Pelechano et al. \cite{PAB08book} and Bu{\c{s}}oniu et al. \cite{BusBab}, respectively.

\subsection{Multi-Agent Navigation}

Numerous models have been proposed to simulate individuals and groups of interacting agents. The seminal
work of Reynolds on \emph{boids} has been influential on this field \cite{R87}. Reynolds used simple local rules to create visually compelling flocks of birds and schools of fishes. Later he extended his model to include autonomous agent behavior \cite{R99}. Since Reynolds's original work, many interesting crowd simulation models have been introduced that account for
groups \cite{BLA03}, cognitive and behavioral rules \cite{FTT99,ST07}, biomechanical principles \cite{GCC+10} and sociological or psychological
factors  \cite{PAB07,GKL+11,popelova2011}. Recent work models the contagion of psychological states in a crowd of agents, for example, in evacuation simulations \cite{tsai2013empirical}. Our approach, in contrast, does not make assumptions about the psychological states of the agents, therefore it is more generally applicable.

An extensive literature also exists on modeling the local dynamics of the agents and computing collision-free
motions. Many different agent-based techniques for collision avoidance have been proposed in control theory~\cite{olfati2006flocking}, traffic simulation~\cite{bham2004high}, animation~\cite{GCK09,KHB+09} and robotics~\cite{Khatib86}. In Section~\ref{sec:bgd} we provide a more detailed description of the collision-avoidance technique we use.

We focus on minimizing the travel time of the agents, but other metrics have been studied in the literature. For example, the work in \cite{rss15,yu2013planning,karamouzas2013space} addresses the problem of minimizing the total length of the path of the agents, formulating the path planning problem as a mixed integer linear program. Coordinating the motion of a set of pebbles in a graph to minimize the number of moves was studied in \cite{kornhauser1984coordinating}.

\subsection{Reinforcement Learning}

Many learning approaches used for robots and agents derive from the reinforcement learning literature \cite{BusBab}. Reinforcement Learning (RL) addresses how autonomous agents can learn by interacting with the environment to achieve their desired goal \cite{Sutton88RL}.
An RL agent performs actions that affect its state and environment, and receives a reward value which
indicates the quality of the performed action.
This reward is used as feedback for the agent to improve its future decisions. Different approaches have been proposed to incorporate RL when multiple agents share the environment (see \cite{BusBab,kober2013,Uther97ARL} for an extensive overview). 

In multi-agent RL algorithms, agents typically need to collect information on how other agents behave and find a policy that maximizes their reward. 
This is expensive when the state space is large and requires a significant degree of exploration to create an accurate model for each agent. Hence, approaches that model the entire environment are focused on small problems and/or a small number of agents. To reduce complexity, some approaches focus on the local interaction neighborhood of each agent \cite{zhang2012coordinated,zhang2013coordinating}. By considering a local neighborhood, the state space of each agent is reduced. To completely avoid the state space complexity, the learning problem can be formulated as a multi-armed bandit problem \cite{Sutton88RL}, where the agents use the reward of each action to make future decisions. In multi-armed bandit problems, the relation between exploiting the current best action and exploring potentially better actions is critical \cite{audibert2009exploration,macready1998bandit}.

\subsubsection{Action Selection Techniques}

A variety of approaches aim at balancing exploration and exploitation, which is critical for online learning problems such as ours. 

A simple approach is $\epsilon$-greedy, which works by selecting the highest valued action with probability 1-$\epsilon$, and a random action with probability $\epsilon$, $0 \leq \epsilon \leq 1$. The value of $\epsilon$ indicates the degree of exploration that the agent performs \cite{SB98}. Because of its probabilistic nature, $\epsilon$-greedy can find the optimal action, 
without being sensitive to the difference between the values of the actions. This means that $\epsilon$-greedy does the same amount of exploration regardless of \emph{how} much better the best known action is, compared to the other actions. 

Another widely used action-selection technique is the upper confidence bounds (UCB) algorithm~\cite{auer2002finite}. UCB is a deterministic method that samples the actions proportionally to the upper-bound of the estimated value of their rewards (based on their current average reward) and their confidence interval (computed using a relation between the number of times each action was selected and the total number of action taken so far by the agent). Unlike $\epsilon$-greedy, UCB considers the value of all actions when deciding which one to choose. However, it does unnecessary exploration when the reward distribution is static (i.e., the best action does not change).

A method that combines the probabilistic nature of $\epsilon$-greedy and that accounts for the changing reward structure is the Softmax action selection strategy. Softmax biases the action choice based on their relative reward value, which means that it increases exploration when all actions have similar value, and it reduces it when some (or one) action is significantly better than the rest. The action selection method we use is based on the Softmax strategy, due to these properties.

\subsection{Learning in Multi-Agent Navigation}

Extensive work has also been done on learning and adapting motion behavior of agents in crowded
environments. Depending on the nature of the learning process, the work can be classified in two main categories: offline and online learning.
In offline learning, agents repeatedly explore the environment and try to learn the optimal policy given an objective function. Examples of desired learned behaviors include collision avoidance, shortest path to destination, and specific group formations.
As an example, the work in~\cite{Henry10} uses inverse reinforcement learning for agents to learn paths from recorded training data.
Similarly, the approach in \cite{Torrey10} applies Q-learning to plan paths for agents in crowds. In this approach, agents learn in a series of episodes the best path to their destination. A SARSA-based \cite{SB98} learning algorithm has also been used in \cite{martinez2012} for offline learning of behaviors in crowd simulations. The approach in \cite{cunningham2012levels} analyzes different strategies for sharing policies between agents to speed up the learning process in crowd simulations. In the area of swarm intelligence, the work in \cite{hettiarachchi2010} uses evolutionary algorithms for robotics,  learning offline the parameters of the fitness function and sharing the learned rules in unknown environments.

Offline learning has significant limitations, which arise from the need to train the agents before the environment is known. In contrast, the main part of our work is an online learning approach. In online approaches, agents are given only partial knowledge of their environment, and are expected to adapt their strategies as they discover more of the environment.
Our approach allows agents to adapt online to unknown environments, and it does not require explicit communication between the agents.

\section{Background}
\label{sec:bgd}

In this section, we first describe different techniques that agents can employ to avoid collisions, specifically focusing on the technique we use in our work.

\subsection{Collision Avoidance}
Methods that have been proposed to prevent collisions during navigation can be classified as \emph{reactive}
and \emph{anticipatory}.

In reactive collision avoidance, agents adapt their motion to other
agents and obstacles along their paths. Many reactive methods \cite{R87,R99,HFV00,Khatib86,Ratering95} use artificial repulsive forces to avoid collisions.
However, these techniques do not anticipate collisions. Only when agents are sufficiently close, they react to avoid collisions. This can lead to oscillations and local minima.
Another limitation of these methods is that the forces must be tuned separately for each scenario, limiting their robustness.

In anticipatory collision avoidance, agents predict and avoid potential upcoming collisions by linearly extrapolating their current velocities. In this line, \emph{geometrically based} algorithms compute collision-free velocities for the agents using either sampling \cite{BLM08,POO+09,KO12,OPOD10} or optimization techniques \cite{Berg09,GCK09}.
\IGNORE{
An important concept in anticipatory collision avoidance approaches is the notion of $velocity$
$obstacle$ (VO) used in robotics \cite{FS98}. For an agent $i$, the VO
represents the set of all the agent's velocities that would eventually result in collision with another agent or obstacle. Van den Berg et al. \cite{BLM08} extended the notion of VO and introduced the concept of $reciprocal$ $velocity$ $obstacle$ (RVO). RVO depends on random sampling of velocities and provides oscillation-free motion between two moving agents. More recently, the principle of \emph{optimal reciprocal collision avoidance} (ORCA) for multi-agent navigation was proposed~\cite{Berg09}. ORCA overcomes the limitations of RVO and guarantees collision-free and oscillation-free motion between moving agents. ORCA accounts for anticipatory avoidance, and reciprocity among the agents as well as sensor noise and motion uncertainty \cite{KGL+13}. An extension to ORCA that accounts for localization uncertainty applied to differential drive robots is presented in \cite{Hennes2012}.
}

\subsection{ORCA}

\begin{figure*}[t]
\centering
\hspace*{\fill}%
\subfloat[Agents \agent{i} and \agent{j} moving at velocities $\mathbf{v_i}$ and $\mathbf{v_j}$, respectively]{
\includegraphics[width=0.35\textwidth]{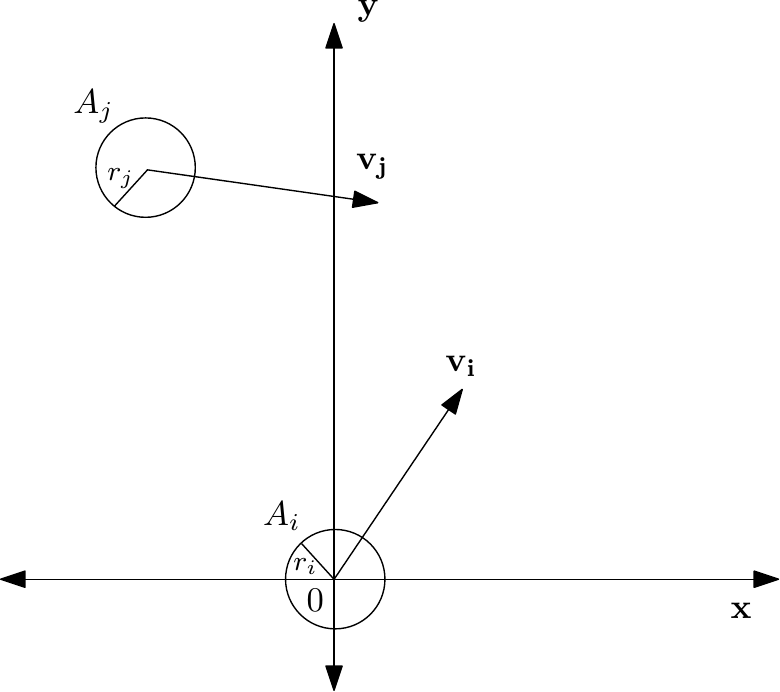}
}\hfill%
\subfloat[\agent{i}'s allowed velocities, in the velocity space]{
\includegraphics[width=0.35\textwidth]{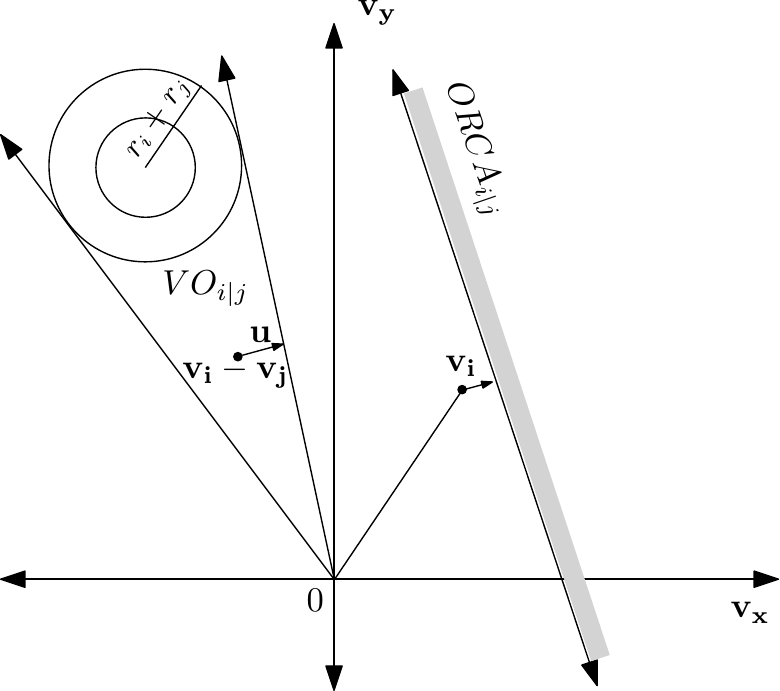}
}
\hspace*{\fill}
\caption{(a) Two agents, \agent{i} and \agent{j}, moving towards a potential collision. (b) The set of allowed velocities for agent $i$  induced by agent $j$  is indicated by the half-plane delimited by the line perpendicular to $\hat{\vv{u}}$ through the point $\mathbf{v_i} + \frac{1}{2}\vv{u}$, where \vv{u} is the vector from $\mathbf{v_i-v_j}$ to the closest point on the boundary of $VO_{i|j}$}
\label{fig:orca}
\end{figure*}

The Optimal Reciprocal Collision Avoidance framework (ORCA) is an anticipatory collision avoidance that builds on the concept of Velocity Obstacles~\cite{FS98}, where agents detect and avoid potential
 collisions by linearly extrapolating their current velocities.
 Given two agents, $A_i$ and $A_j$, the set of velocity
obstacles $VO_{A_i|A_j}$ represents the set of all relative velocities between $i$  and $j$  that will
result in a collision at some future moment. Using the VO formulation, we can guarantee collision avoidance
by choosing a relative velocity that lies outside the set $VO_{A_i|A_j}$. Let $\vv{u}$ denote the minimum
change in the relative velocity of $i$  and $j$  needed to avoid the collision. ORCA assumes that the two
agents will \emph{share} the responsibility of avoiding it and requires each agent to change its current velocity by at least $\frac{1}{2}\vv{u}$. Then, the set of feasible velocities for $i$  induced by $j$ is the half-plane of velocities given by:
\begin{align*}
ORCA_{A_i|A_j} = \{\vel{}\; | (\vel{}-(\vel{i} + \frac{1}{2}\vv{u}))\cdot\hat{\vv{u}}\},
\end{align*}
where $\hat{\vv{u}}$ is the normalized vector $\vv{u}$  (see Fig.~\ref{fig:orca}).
Similar formulation can be derived for determining \agent{i}'s permitted velocities with respect to a static obstacle $O_k$. We denote this set as $ORCA_{A_i|O_k}$.

The overall approach works as follows. At each time step of the simulation, each agent $i$ uses its goal-oriented velocity $\vgoal{i}$ and computes a new collision-free velocity
by taking into account its neighboring agents and static obstacles.
First, agent $i$ infers its set of \emph{feasible} velocities, $FV_{A_i}$, from the intersection of all permitted half-planes $ORCA_{A_i|A_j}$ and $ORCA_{A_i|O_k}$ induced by each neighboring agent $j$ and obstacle $O_k$, respectively. Having computed $FV_{A_i}$, the agent selects a new velocity \vnew{i} for itself that is closest to its preferred velocity $\vpref{i}$ and lies inside the region of feasible velocities:
\begin{equation}
\label{eq:vnew}
\vnew{i} = \argmin_{\vel{} \in FV_{A_i}} \|\vel{} - \vpref{i} \|.
\end{equation}
The optimization problem in \eqref{eq:vnew} can be efficiently solved using linear programming, since $FV_{A_i}$ is a convex region bounded by linear constraints.
Finally, agent $i$ updates its position based on the newly computed velocity. As ORCA is a decentralized approach, each agent computes its velocity independently.

In addition, each agent typically uses its goal-oriented velocity $\vgoal{i}$ as an input preferred velocity to ORCA in \eqref{eq:vnew}.

\subsection{Limitations of ORCA}

\begin{figure*}[t]
\centering
\subfloat[Start positions]{\includegraphics[width=.24\textwidth]{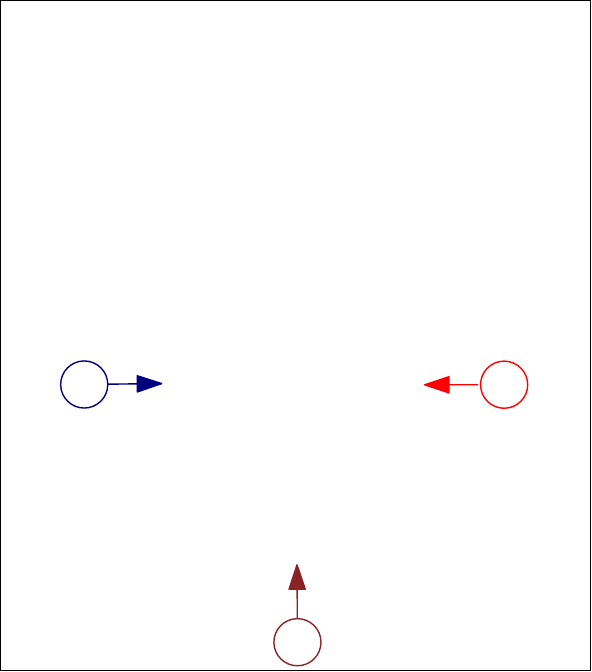}} 
\subfloat[Goal positions]{\includegraphics[width=.24\textwidth]{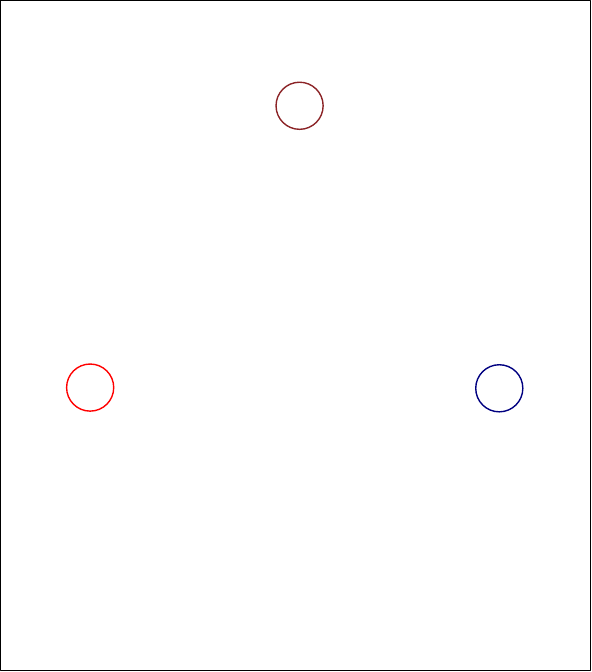}}
\subfloat[ORCA]{\includegraphics[width=.24\textwidth]{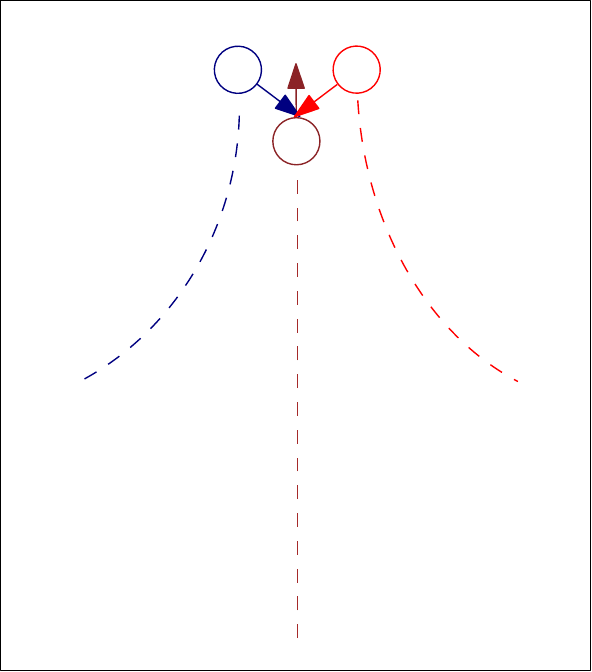}}
\subfloat[ALAN]{\includegraphics[width=.24\textwidth]{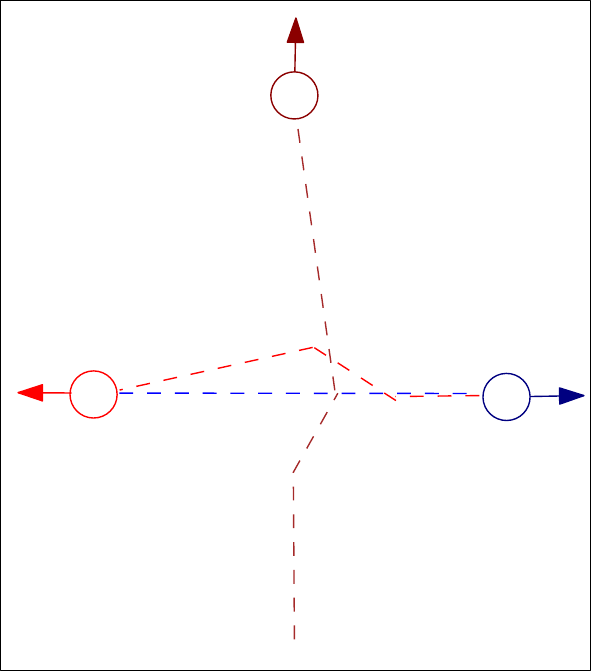}}
\caption{Three agents cross paths.  (a) Initial positions of the agents. (b) Goal positions of the agents. (c) When navigating with ORCA, the agents run into and push each other resulting in inefficient paths. (d) When using ALAN the agents select different preferred velocities which avoid local minima, resulting in more efficient paths.}
\label{fig:3agentpos}
\end{figure*}

Although ORCA guarantees collision-free motions and provides a locally optimal behavior for each agent, the lack of coordination between agents can lead to globally inefficient motions. For an example, see Fig.~\ref{fig:3agentpos}. Here, three
agents start from the initial positions (Fig.~\ref{fig:3agentpos})(a) and
must reach the final positions (Fig.~\ref{fig:3agentpos})(b). Because
the agents follow only their goal-oriented preferred velocity, they will get stuck in a local minimum which generates the trajectories shown in Fig.~\ref{fig:3agentpos}(c). If instead the agents behaved differently, for instance, by selecting a different \vpref{} for a short period of time, they might find a larger region of feasible velocities. This might indirectly help avoid/solve the overall congestion, benefiting all agents. Our proposed approach, ALAN, directly addresses this limitation, allowing agents to adapt their preferred velocity online, hence improving their motion efficiency. An example of the trajectories generated by our approach can be seen in Fig.~\ref{fig:3agentpos}(d).

\section{Problem Formulation}
\label{sec:man}

In our problem setting, given an environment and a set $A$ of agents, each with a start and a goal 
position, our goal is 
to enable the agents to reach their goals as soon as possible and without collisions. We also require that the agents move
\emph{independently} 
and without explicitly communicating with each other. For simplicity, we model each agent as a disc which moves on a 2D plane that may also contain a set of $k$ static obstacles $O$ (approximated by line segments in all our experiments).

Given $n$ agents,
let agent $A_i$ have
radius $r_i$, goal position $\mathbf{g}_{i}$, and maximum speed $\upsilon_i^{\max}$.
Let also $\mathbf{p}_i^t$ and $\mathbf{v}_i^t$ denote the agent's position and velocity, respectively, at time $t$.
Furthermore, agent $A_i$ has a preferred velocity $\vpref{i}$ at which it prefers to move.
Let $\vgoal{i}$ be the preferred velocity directed towards the agent's goal $\mathbf{g}_{i}$ with a magnitude equal to $\upsilon_i^{\max}$.
The main objective of our work is to minimize the travel time of the set of agents $A$ to their goals, while guaranteeing collision-free motions. To measure this global travel time, we could consider the travel time of the last agent that reaches its goal. However, this value does not provide any information of the travel time of all the other agents. Instead,  we measure this travel time, $TTime(A)$, as a linear combination of the average travel time of all the agents in $A$ and its spread. Formally:
\begin{equation}
\label{eq:ttime}
\begin{array}{lll}
TTime(A) &=& \mu\left(
TimeToGoal(A)\right) \\
&& +~ 3\ \sigma\left(
TimeToGoal(A)\right)
\end{array}
\end{equation}

\noindent where $TimeToGoal(A)$ is the set of travel times of all agents in $A$ from their start positions to their goals, and $\mu(\cdot)$ and $\sigma(\cdot)$ are the 
average and the standard deviation (using the unbiased estimator) of the set $TimeToGoal(A)$, respectively.
If the times to goals of the agents follow a normal distribution, then
$TTime(A)$ represents the upper bound of the $TimeToGoal(A)$ for approximately 99.7\% of the agents.  Even if the distribution is not normal, at least 89\% of the times will fall within three standard deviation (Chebyshev's inequality).  Our objective can be formalized as follows:
\begin{equation}
\label{eq:globaltime}
\begin{aligned}
& \text{minimize} &&  TTime(A)\\
& \text{s.t.} &&  \|\mathbf{p}_{i}^t-\mathbf{p}_{j}^t\| > r_{i}+ r_{j}, \underset{i \neq j}\forall  i, j \in [1,n] 
\\
&
& & dist(\mathbf{p}_{i}^t, O_j) > r_{i},  \forall  i \in [1,n], j\in [1,k] 
\\
&
& & \|\mathbf{v}_{i}^t\|  \leq \upsilon_i^{\max},\;\;\ \ \ \ \ \forall i \in [1,n]\;\ 
\end{aligned}
\end{equation} 
\noindent where $dist(\cdot)$ denotes the shortest distance between two positions.
To simplify the
notation, in the rest of the paper we omit the index of the specific agent being referred, unless it is needed for clarity.

Minimizing Eq.~\ref{eq:globaltime} for a large number of agents using a centralized planner with complete information is intractable
(PSPACE-hard~\cite{hopcroft1984complexity}), given the combinatorial nature of the optimization problem and the continuous space of movement for the agents. Since we require that the agents navigate \emph{independently} and without explicit communication with each other, Eq.~\ref{eq:globaltime} has to be minimized in a decentralized manner. As the agents do not know in advance which trajectories are feasible, the problem becomes for each agent to decide how to move at each timestep, given its perception of the local environment.
This is the question addressed by our online learning approach, ALAN, which is described next.

\section{ALAN}
\label{sec:mainalan}

ALAN is composed by an action selection framework, which provides a set of preferred velocities an agent can choose from, and a reward function the agent uses to evaluate the velocities and select the velocity to be used next. ALAN keeps an updated reward value for each action using a moving time window of the recently obtained rewards.  
If information about the set of navigation environments is available, ALAN can take advantage of an \textit{action learning} approach to compute, in an offline manner, an action set that is optimized for one or a set of scenarios (see Section \ref{sec:mcmc}).

In ALAN, each agent runs a continuous cycle of sensing and action until it reaches its destination. To guarantee real-time behavior, we impose a hard time constraint of 50 ms per cycle. 
We assume that the radii, positions and velocities of nearby agents and obstacles can be obtained by sensing.
At each cycle the agent senses and computes its new collision-free velocity which is used
until the next cycle. The 
velocity has to respect the agent's geometric and kinematics constraints while ensuring progress towards its goal.

To achieve this, ALAN follows a two-step process. First, the agent selects a preferred velocity \vpref{} (as described later in  
Section \ref{sec:acsel}). Next, this \vpref{} is passed to ORCA which produces a collision-free velocity \vnew{}, which is the velocity the agent will use during the next timestep. 

Algorithm~\ref{algo:actionsel} shows an overview of ALAN. This algorithm is executed at every cycle. If an action is to be selected in the current cycle (line 3, in average every 0.2 secs.), the Softmax action selection method (presented later in Section \ref{sec:acsel}) returns a \vpref{} (line 4), which is passed to ORCA. After computing potential collisions, ORCA returns a new collision-free velocity \vnew{} (line 6), and the $getAction$ method returns the ID of the action $a$ that corresponds to the \vpref{} selected (line 7). This action $a$ is executed (line 8), which moves the agent with the collision-free velocity \vnew{} for the duration of the cycle, before updating the agent's position for the next simulation step (line 9). 
The agent determines the quality of the action $a$ (lines 10-12) by computing its reward value (see Section \ref{sec:rw}). 
This value becomes available to the action selection mechanism, which will select a new \vpref{} in the next cycle. This cycle repeats until the agent reaches its goal.

\begin{algorithm}[!ht]
\caption{The ALAN algorithm for an agent}
\label{algo:actionsel}
\begin{algorithmic}[1]
\STATE initialize simulation
\WHILE {not at the goal}  %
	\IF {$UpdateAction(t)$}
       	\STATE $\vpref{} \gets Softmax(Act)$
	\ENDIF
    \STATE \emph{$\vnew{} \gets ORCA(\vpref{})$}
    \STATE $a \gets getAction(\vpref{})$
    \STATE $Execute(a)$
	\STATE \emph{$\post{} \gets \postt{} + \vnew{} \cdot \Delta{t}$}
	\STATE \emph{$\mathcal{R}^{goal}_a \gets GoalReward(a^{t-1})$}  (cf. Eq.~\ref{eq:goalmotion})
	\STATE \emph{$\mathcal{R}^{polite}_a \gets PoliteReward(a^{t-1})$} (cf. Eq.~\ref{eq:coopmotion})
	\STATE \emph{$\mathcal{R}_a \gets (1-\gamma) \cdot \mathcal{R}^{goal}_a + \gamma \cdot \mathcal{R}^{polite}_a$}
\ENDWHILE
\end{algorithmic}
\end{algorithm}

\begin{figure}[!ht]
\centering
\includegraphics[width=.6\columnwidth]{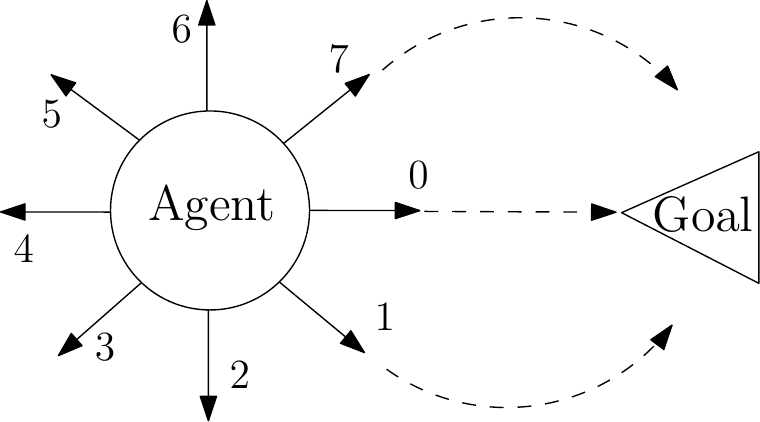}\; \;\; \; \; \;
 \caption{Example set of actions with the corresponding action ID. The eight actions correspond to moving at 1.5 $m/s$ with different angles with respect to the goal: $0^\circ$, $45^\circ$, $90^\circ$, $135^\circ$, $-45^\circ$, $-90^\circ$, $-135^\circ$ and $180^\circ$.}
\label{fig:actions15}
\end{figure}

The main issue is how an agent should choose its preferred velocity. 
Typically, an agent would prefer a velocity that drives it closer to its goal, but different velocities may help the entire set of agents to reach their destinations faster (consider, for example, an agent moving backwards to alleviate congestion). Therefore, we allow the agents to use different \emph{actions}, which correspond to different preferred velocities (throughout the rest of this paper, we will use the terms preferred velocities and actions interchangeably).
In principle, finding the best motion would require each agent to make a choice at every step in a continuous 2D space, the space of all possible speeds and directions. This is not practical in real-time domains. 
Instead, agents plan their motions over a discretized set of a small number of preferred velocities, the set $Act$. 
An example set of 8 actions uniformly distributed in the space of directions is shown in Figure~\ref{fig:actions15}. We call this set $Sample$ set. 

Different action sets affect the performance of the agents. We analyze this later (Section~\ref{sec:mcmc}), where we present an offline learning method to find an optimal set of actions. 

\subsection{Reward Function}
\label{sec:rw}

\begin{figure*}[!ht]
 \centering
 \includegraphics[width=0.75\textwidth]{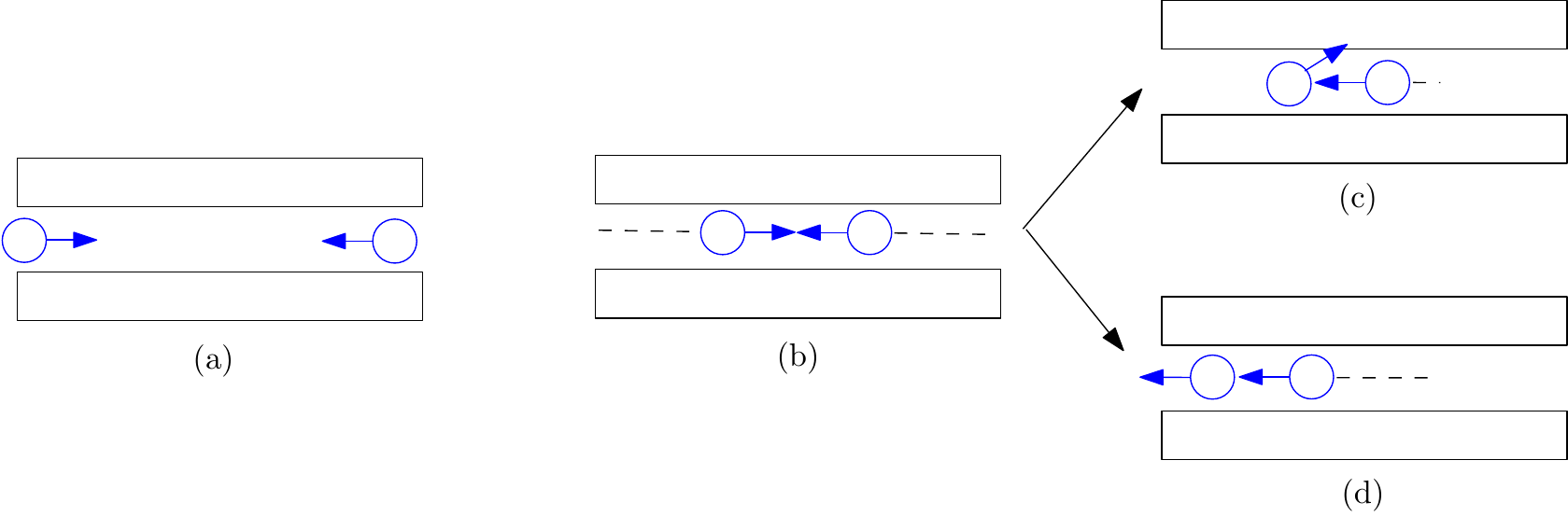}
 \caption{Two agents moving to their goals in opposite sides of the corridor. Different behaviors are produced by optimizing different metrics.  (b) When meeting in the middle of the corridor, agents cannot continue their goal motions without colliding. (c) Considering only goal progress when choosing actions results in one agent pushing the other out of the corridor. (d) Considering both goal progress and effect of action in other agents results in one agent moving backwards to help the other move to its goal.}
 \label{fig:exreward}
\end{figure*}

The quality of an agent's selected action \vpref{} is evaluated based on two criteria: how much it moves the agent to its goal, and its effect on the motion of nearby agents. The first criterion allows agents to reach their goals, finding non-direct goal paths when facing congestion or static obstacles. The second criterion encourages actions that do not slow down the motion of other agents. To do this, agents take advantage of the reciprocity assumption of ORCA: when a collision is predicted, both potentially colliding agents will deviate to avoid each other. Hence, if a collision-free \vnew{} computed by ORCA is significantly different from the selected preferred velocity \vpref{}, it represents a deviation of the same magnitude for another agent. Therefore, to minimize the negative impact of its decisions on the nearby agents, i.e., to be \emph{polite} towards them, each agent should choose actions whose \vnew{} is similar to the \vpref{} that produced it. This duality of goal oriented and ``socially aware" behaviors, in humans, has been recently studied in \cite{sieben2017collective}.

Observe the example navigation task in Figure \ref{fig:exreward} where two agents must travel to the other side of a small corridor. In Figure \ref{fig:exreward}(c), agents use only goal progress as a criterion for evaluating their actions. In Figure \ref{fig:exreward}(d), they use both goal progress and the effect in the motion of other agents as criteria for evaluating their actions. Using the $Sample$ action set defined in Fig.~\ref{fig:actions15}, agents first move to the center of the corridor. If agents only optimize their goal progress, eventually one of them will start pushing the other out of the corridor as they explore their actions (Figure \ref{fig:exreward}(c)). This pushing behavior continues slowly until the agents exit the corridor and find space to avoid each other and eventually reach their goals. 
If agents also consider the effect of their actions on others, then one of them will eventually find that the backwards action does not add constraints to the motion of the other agent, hence can help that agent to move to its goal. The agent then willingly moves backwards (Figure \ref{fig:exreward}(d)), exiting the corridor faster and reducing the travel time for both agents. 
We show that considering both criteria in the evaluation of each action reduces the travel time of the agents overall.

Specifically, we define the reward $\mathcal{R}_a$ for an agent performing action $a$ to be a convex combination of a \emph{goal-oriented} component
and a \emph{politeness} component:
\begin{equation} \label{eq:reward}
\mathcal{R}_a = (1-\gamma) \cdot \mathcal{R}^{goal}_a + \gamma \cdot \mathcal{R}^{polite}_a,
\end{equation}
where the parameter $\gamma$, called $coordination$ $factor$, controls the influence of each component in the total reward (0 $\leq \gamma <$ 1).

The \emph{goal-oriented} component $\mathcal{R}^{goal}_a$
computes the scalar product of the collision-free velocity \vnew{} of the agent with the normalized vector pointing from the
position \pos{} of the agent to its goal \goal{}. This component promotes preferred velocities that lead the agent as quickly as possible to its goal.
More formally:
\begin{equation} \label{eq:goalmotion}
\mathcal{R}^{goal}_a = \vnew{} \cdot  \frac{\goal{} - \pos{}}{\|\goal{}-\pos{}\|}
\end{equation}

The \emph{politeness} component $\mathcal{R}^{polite}_a$ compares the executed preferred velocity with the resulting collision-free velocity.
These two velocities will be similar when the preferred velocity does not conflict with other agents' motions, and will be different when it leads to potential collisions. 
Hence, the similarity between \vnew{} and \vpref{} indicates how polite is the corresponding action, with respect to the motion of the other agents.
Polite actions reduce the constraints on other agents' motions, allowing them to move and therefore advancing the global simulation state.
Formally:
\begin{equation} \label{eq:coopmotion}
\mathcal{R}^{polite}_a = \vnew{} \cdot \vpref{}
\end{equation}

If an agent maximizes $\mathcal{R}^{goal}_a$, it would not consider the effects of its actions on the other agents.
On the other hand, if the agent tries to maximize $\mathcal{R}^{polite}_a$,
it has no incentive to move towards its goal, which means it might never reach
it. Therefore, an agent should aim at maximizing a combination of both components. Different behaviors may be obtained with different values of $\gamma$.
In Section~\ref{sec:gamma}, we analyze how sensitive the performance of ALAN is to different values of $\gamma$. Overall, we  found that $\gamma = 0.4$ provides an appropriate balance between these two extremes.
\begin{figure}[!ht]
 \centering
 \includegraphics[width=0.75\columnwidth]{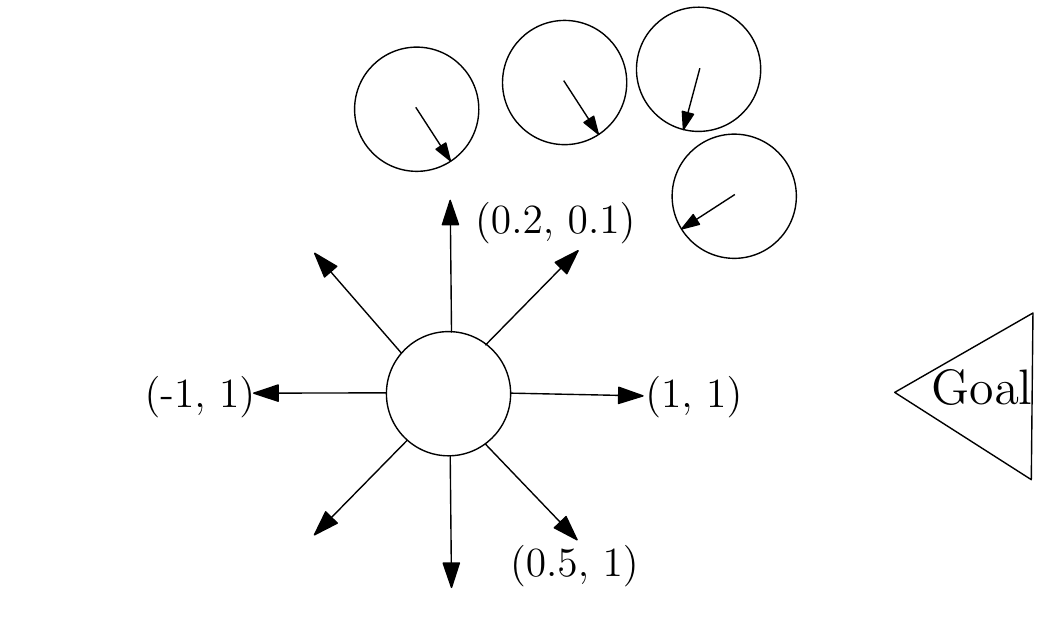}
 \caption{Example of reward values for different actions under clear and congested local conditions. The reward $\mathcal{R}_a$ of each action $a$ is shown as a pair of goal-oriented and a politeness components ($\mathcal{R}^{goal}_a$, $\mathcal{R}^{polite}_a$).}
 \label{fig:actions2}
\end{figure}

Figure~\ref{fig:actions2} shows an example of conditions an agent may encounter.
Here, there is congestion on one side of the agent, which results in low
reward values for the left angled motion. The other actions are not constrained, and consequently their reward value is higher. In this case, the agent will choose the straight goal-oriented action,
as it maximizes $\mathcal{R}_a$.

\subsection{Multi-armed Bandit Formulation}

As the number of navigation states is very large, we adapt a stateless representation. Each agent can select one action at a time, hence the question is which one should the agent execute at a given time. 
In ALAN, agents learn the reward value of each action through its execution, in an online manner, and keep the recently obtained rewards (using a moving time window of the rewards) to decide how to act. We allow a chosen action to be executed for a number of cycles, and perform an a-posteriori evaluation to account for bad decisions. This way, the problem of deciding how to move becomes a resource allocation problem, where agents have a set of alternatives strategies and have to learn their estimated value via sampling, choosing one at each time in an online manner until reaching their goals.

Online learning problems with a discrete set of actions and stateless representation can be well formulated as multi-armed bandit problems. In a multi-armed bandit problem, an agent makes sequential decisions on a set of actions to maximize its expected reward.  This formulation is well-suited for stationary problems, as existing algorithms guarantee a logarithmic bound on the regret. Although our problem is non-stationary in a global sense, as the joint local conditions of the agents are highly dynamic, individual agents often undergo long periods of stationary reward distributions. 

\begin{figure}[!ht]
 \centering
 \includegraphics[width=0.6\columnwidth]{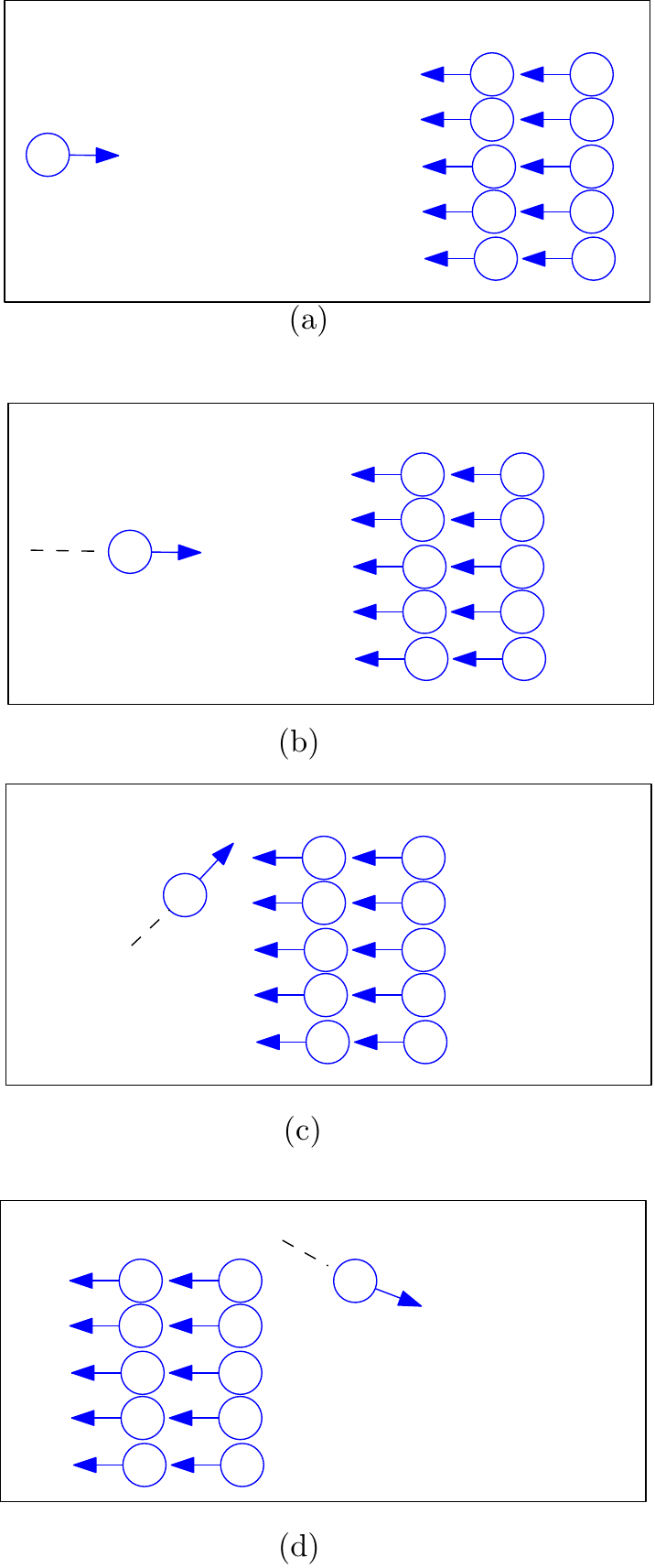}
 \caption{Distinguishable periods of stationary reward distribution for the agent on the left. (a) The agent must reach its goal on the other side of a group of agents moving in the opposite direction. The optimal action in each period changes between (b) the goal oriented motion, (c) the sideways motion to avoid the incoming group, and (d) the goal oriented motion again, once the agent has avoided the group.}
 \label{fig:stationary}
\end{figure}

An example of a solution for a navigation task, where we can distinguish periods of stationary reward distributions, is shown in Figure \ref{fig:stationary}. Here, a single agent on the left must travel to the other side of an incoming group of agents (Fig. \ref{fig:stationary}(a)). Initially, the optimal action for the single agent is to directly move towards its goal, as it has not yet sensed the incoming group (Fig. \ref{fig:stationary}(b)). The optimal action does not change until the agent sees the group of agents in its direct goal path, at which point the best action is to move sideways towards the goal (Fig. \ref{fig:stationary}(c)) to avoid the group. This sideways motion is the locally optimal action until the agent has reached the clear area. 
Finally, once the agent has avoided the group, the locally optimal action is again to move straight towards the goal (Fig. \ref{fig:stationary}(d)). Hence, in this example we see three periods of stationary reward distribution.

Therefore, by learning the action that maximizes a local reward function (Eq.~\ref{eq:reward}) in each of these stationary periods, agents can adapt to the local conditions.

\subsection{Action Selection}
\label{sec:acsel}

We now describe how ALAN selects, at each action decision step, one of the available actions based on their computed reward values and a probabilistic action-selection strategy, Softmax, which is described next.

\subsubsection{Softmax}
\label{sec:softmax}

Softmax is a general action selection method that balances exploration and exploitation in a probabilistic manner~\cite{SB98,ziebart2009planning,whiteson2007empirical}. 
This method biases the action selection towards actions that have higher value (or reward, in our terminology), by making the probability of selecting an action dependent on its current estimated value. The most popular Softmax method uses the Boltzmann distribution to select among the actions. Assuming that $\mathcal{R}_a$ is the reward value of action $a$, the probability of choosing $a$ is given by the following equation:
\begin{equation} \label{eq:softmax}
Softmax(a) =  {\exp\left({\frac{\mathcal{R}_a}{\tau}}\right) } \left/{\sum_{a=1}^{|Act|}\exp\left({\frac{\mathcal{R}_a}{\tau}}\right) }\right.
\end{equation}
The degree of exploration performed by a Boltzmann-based Softmax method is controlled by the parameter $\tau$, also called the \emph{temperature}. With values of $\tau$ close to zero the highest-valued actions are more likely to be chosen, while high values of $\tau$ make the probability of choosing each action similar. We use a value of $\tau$=0.2, as we found that it shows enough differentiation between different action values without being too greedy.

Another critical design issue of our action selection method is the duration of the time window used. 
Keeping old samples with low values might make a good action look bad, but discarding them too quickly will ignore the past. Because of this, we use a moving time window of the most recently obtained rewards, and compute the estimated value of each action based only on the rewards in that time window, using the last sampled reward for each. If an action has not been sampled recently, it is assumed to have a neutral (zero) value, which represents the uncertainty of the agent with respect to the real value of the action. Actions with a neutral value have a low probability of being selected if the currently chosen action has a ``good" value ($>$0), and have a high probability of being selected if the currently chosen action has a ``bad" value ($<$0). When making an action decision, an agent retrieves the last sampled reward value for each action in the time window, or zero if the action has not been sampled recently. These values are then used by Softmax (Eq. \ref{eq:softmax}) to determine the probability of each action being chosen. 

In Section~\ref{sec:twresult} we analyze the effect of different sizes of time window on the performance of ALAN. 

\begin{figure*}[!ht]
 \centering
 \includegraphics[width=0.8\textwidth]{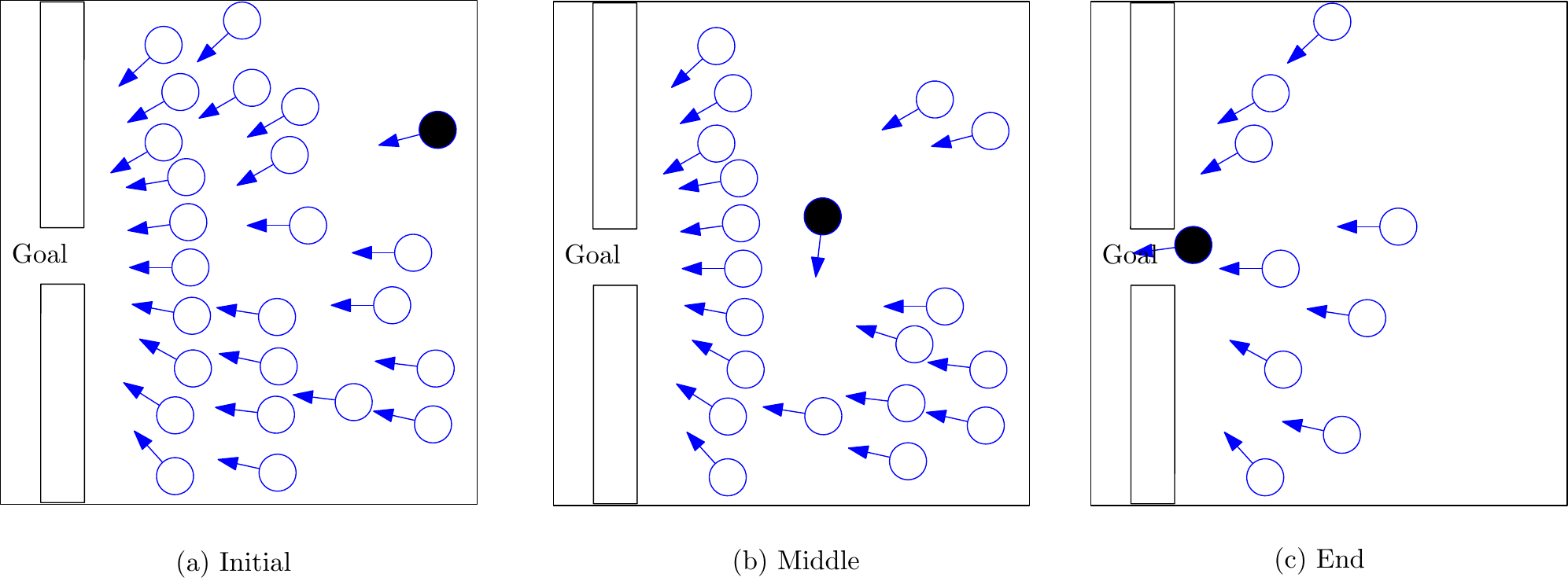}
 \caption{Screen shots of three states of a navigation problem. (a) Initially, the black agent can move unconstrained towards the goal. (b) During its interaction with other agents, the black agent moves sideways since this increases its reward. (c) Finally, when its goal path is free, the black agent moves again towards the goal.}
 \label{fig:states}
\end{figure*}

\begin{table*}[t]
	\centering
    \begin{tabular}{ | l | l| r | r | r | r | r | r | r | r |}
    \hline
    \multicolumn{2}{|c|}{\multirow{2}{*}{Simulation state}}  & \multicolumn{8}{|c|}{Action ID} \\ \cline{3-10}
 \multicolumn{2}{|c|}{}  & 0 & 1 & 2 & 3 & 4 & 5 & 6 & 7 \\ \hline
   \multirow{2}{*}{Initial} & reward & \textbf{0.997} & 0 & 0 & 0.147 & 0 & 0.145 & 0 & 0   \\ 
    & prob & \textbf{94.1}\% & 0.64\% & 0.64\% & 
    1.34\% & 0.64\% & 1.33\% & 0.64\% & 0.64\%   \\ \hline
   \multirow{2}{*}{Middle} 
      & reward &-0.05 & -0.42 & -0.54 & 0 & 0.001 & -0.192 & \textbf{0.456} & 0 \\ 
      & prob & 5.4\% & 0.83\% & 0.46\% & 7.1\% & 7.1\% & 2.7\% & \textbf{69.3}\% & 7.1\% \\ \hline
    \multirow{2}{*}{End} 
      & reward & \textbf{0.63} & 0.47 & 0  & 0.48 & 0 & 0 & 0.177 & 0\\
      & prob & \textbf{56.7}\% & 25\% & 2.4\%  & 3\% & 2.4\% & 2.4\% & 5.8\% & 2.4\%\\ \hline
    \end{tabular}
    \caption{Reward values and probability for each action of being chosen by the black agent using ALAN in the three different states shown in Figure \ref{fig:states}.}
    \label{tab:rewards-prob}
\end{table*}

\subsubsection{Evolution of rewards during simulation}
As agents move to their goals, their evaluation of the available actions affects the probability of choosing each action. Figure \ref{fig:states} shows three simulation states of a navigation task while Table \ref{tab:rewards-prob} shows, for each action of the black agent, the computed rewards and probability of being chosen as the next action. The goal of this evaluation is to empirically show how the estimated value of each action changes as the agent faces different conditions, and how these estimates affect the probability of the action being chosen.

In the Initial state (Fig. \ref{fig:states}(a)), the black agent can move unconstrained towards the goal, which is reflected in the high reward and corresponding probability of the goal oriented action (ID 0). In the Middle state (Fig. \ref{fig:states}(b)), the black agent is facing congestion, which translates into a low reward for the goal oriented action. Instead, it  determines that the action with the highest value is the one moving left (ID 6), which also has the highest probability of being chosen. Finally, in the End state (Fig. \ref{fig:states}(c)), the goal path of the black agent is free. Through exploration, the black agent determines that the goal oriented motion (ID 0) is again the one with the best value, though with lower reward value than in the beginning, as the wall prevents the agent from moving at full speed. With a 56.7\% probability, the agent selects the goal oriented motion and eventually reaches its goal. Note that the actions not sampled during the duration of the time window used in this experiment (2s) are assigned the neutral zero value.

\section{Evaluation}
\label{sec:quant}
 
We now present the experimental setup, performance metrics, and scenarios used to compare the performance of ALAN to other navigation approaches (Section \ref{sec:comparisons}). We also evaluate the design choices of ALAN, such as the action selection method (Section \ref{sec:method}), the time window length (Section \ref{sec:twresult}), and the balance between goal progress and politeness, controlled by the coordination factor $\gamma$ (Section \ref{sec:gamma}) in the reward function. 
Additional results are presented later, after we extend the action selection method to include learning the action space.
 
\subsection{Experimental Setup}
\label{sec:exp}

We implemented ALAN in C\verb!++!. Results were gathered on an Intel Core i7 at 3.5 GHz. Each experimental result is the average over 30 simulations. In all our runs, we updated the positions of the agents every $\Delta t=50\,$ms and set the maximum speed \umax{} of each agent to $1.5\,$m/s and its radius \radius{} to $0.5\,$m. Agents could sense other agents within a $15\,$m radius, and obstacles within $1\,$m. To avoid synchronization artifacts, agents are given a small random delay in how frequently they can update their \vpref{} (with new \vpref{} decisions computed every $0.2\,$s on average). This delay also gives ORCA a few timesteps to incorporate sudden velocity changes before the actions are evaluated. Small random perturbations were added to the preferred velocities of the agents to prevent symmetry problems.

\begin{figure*}[t]
 \centering
 \includegraphics[width=0.8\linewidth]
 {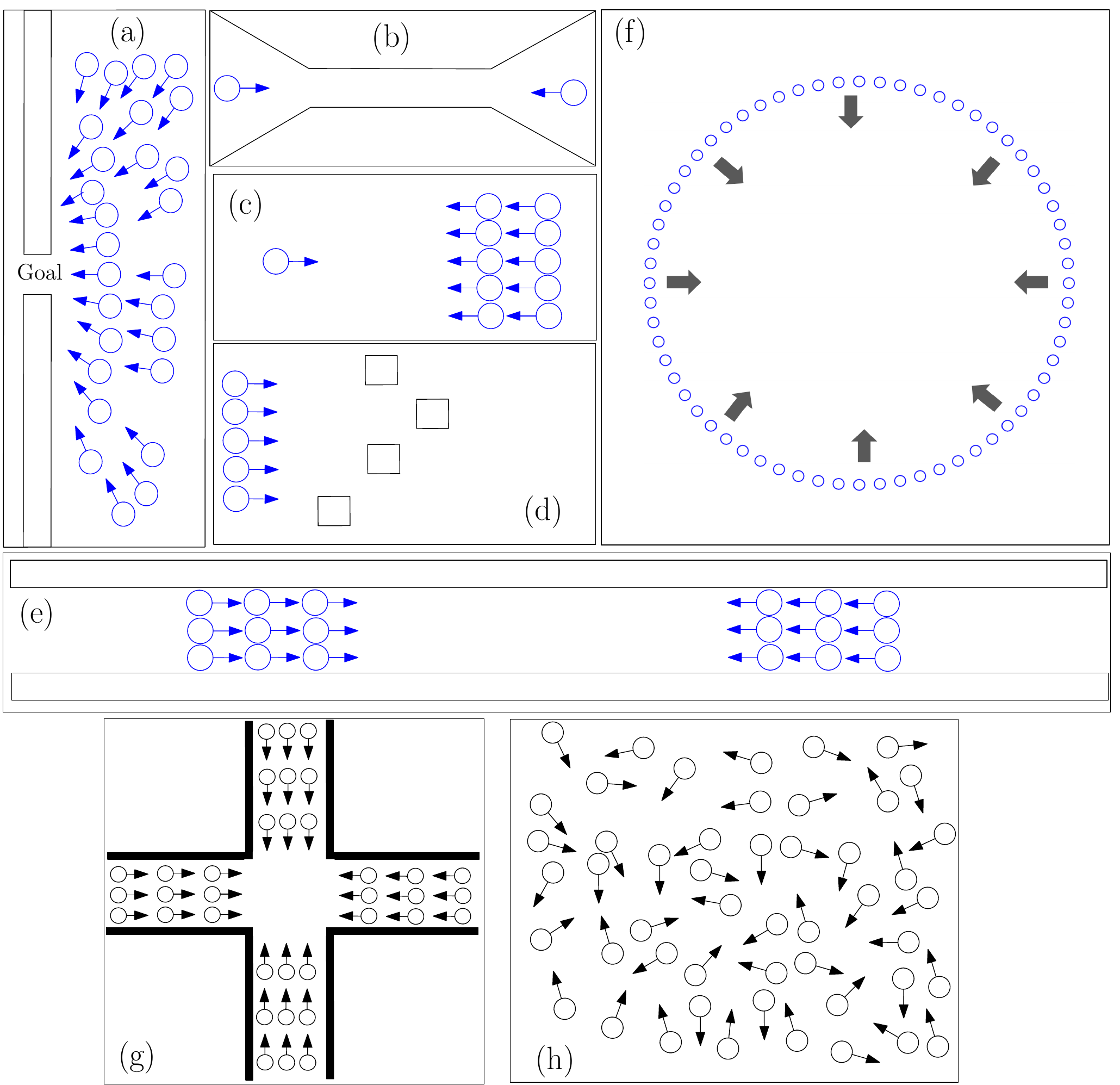}
 \caption{Simulated scenarios:(a) \textsc{Congested}, (b) \textsc{Deadlock}, (c) \textsc{Incoming}, (d) \textsc{Blocks}, (e) \textsc{Bidirectional}, (f) \textsc{Circle}, (g) \textsc{Intersection} and (h) \textsc{Crowd}.}
 \label{fig:scenarios}
\end{figure*}

\subsection{Performance Metric}
\label{sec:io}
To evaluate the performance of ALAN, we measure the time that the agents take to reach their goals compared to the upper bound of their theoretical minimum travel time. 
We call this metric interaction overhead.\\

\noindent\emph{Definition: Interaction Overhead. } 
The interaction overhead is the difference between the travel time of the set of agents, as measured by Eq.~\ref{eq:ttime}, and the upper bound of their travel time if all the agents could follow their shortest paths to their goals at maximum speed without interacting with each other, i.e.:
\begin{equation*} \label{eq:overhead2}
{Interaction\ Overhead}= TTime(A) - MinTTime(A)\\
\end{equation*} 
\noindent where $MinTTime(A)$ is the upper bound of the theoretical minimum travel time of the set of agents $A$, computed similarly to Eq.~\ref{eq:ttime}, and evaluated as follows:

\begin{multline}
MinTTime(A) = \mu\left(MinimumGoalTime(A))
\right)\\ 
+ 3 \sigma\left(MinimumGoalTime(A) \right)
\end{multline} 

\noindent where $MinimumGoalTime(A)$ is the set of travel times for all agents in $A$, if they could follow their shortest route to their goals, unconstrained, at maximum speed.

The interaction overhead metric allows us to evaluate the performance of ALAN from a theoretical standpoint in each of the navigation scenarios.

An interaction overhead of zero represents a lower bound on the optimal travel time for the agents, and it is the best result that any optimal centralized approach could potentially achieve.

\subsection{Scenarios}

To evaluate ALAN we used a variety of scenarios, with different numbers of agents and, in some cases, with static obstacles.
Figure~\ref{fig:scenarios} shows the different simulation scenarios. These include: (a) \textsc{Congested}: 32 agents are placed very close to the narrow exit of an open hallway and must escape the hallway through this exit (Fig.~\ref{fig:scenarios}(a)); (b) \textsc{Deadlock}: Ten agents start at opposite sides of a long, narrow corridor. Only one agent can fit in the narrow space (Fig.~\ref{fig:scenarios}(b)); (c) \textsc{Incoming}: A single agent interacts with a group of 15 agents moving in the opposite direction (Fig.~\ref{fig:scenarios}(c)); (d) \textsc{Blocks}: Five agents must avoid a set of block-shaped obstacles to reach their goals (Fig.~\ref{fig:scenarios}(d)); (e) \textsc{Bidirectional}: two groups of 9 agents each move in opposite directions inside a corridor (Fig.~\ref{fig:scenarios}(e));  (f) \textsc{Circle}: 80 agents walk to their antipodal points on a circle  (Fig~\ref{fig:scenarios}(f));  (g) \textsc{Intersection}: 80 agents in four perpendicular streams meet in an intersection (Fig~\ref{fig:scenarios}(g));  (h) \textsc{Crowd}: 400 randomly placed agents must reach their randomly assigned goal positions, while moving inside a squared room (Fig~\ref{fig:scenarios}(h)).

\subsection{Comparison of ALAN to Other Navigation
Approaches}
\label{sec:comparisons}

To better quantify the effect of ALAN in reducing the travel time of the agents, we compare its interaction overhead values with existing navigation algorithms: ORCA, the Social Forces model proposed by Helbing et al.~\cite{helbing1995social}, that has been extensively used to simulate the navigation of pedestrians~\cite{HFV00,johansson2007specification,HBW03,helbing2001self}, and the Predictive collision avoidance model proposed in \cite{KHB+09}.

\begin{figure}[!ht]
 \centering
 \includegraphics[width=0.99\columnwidth]{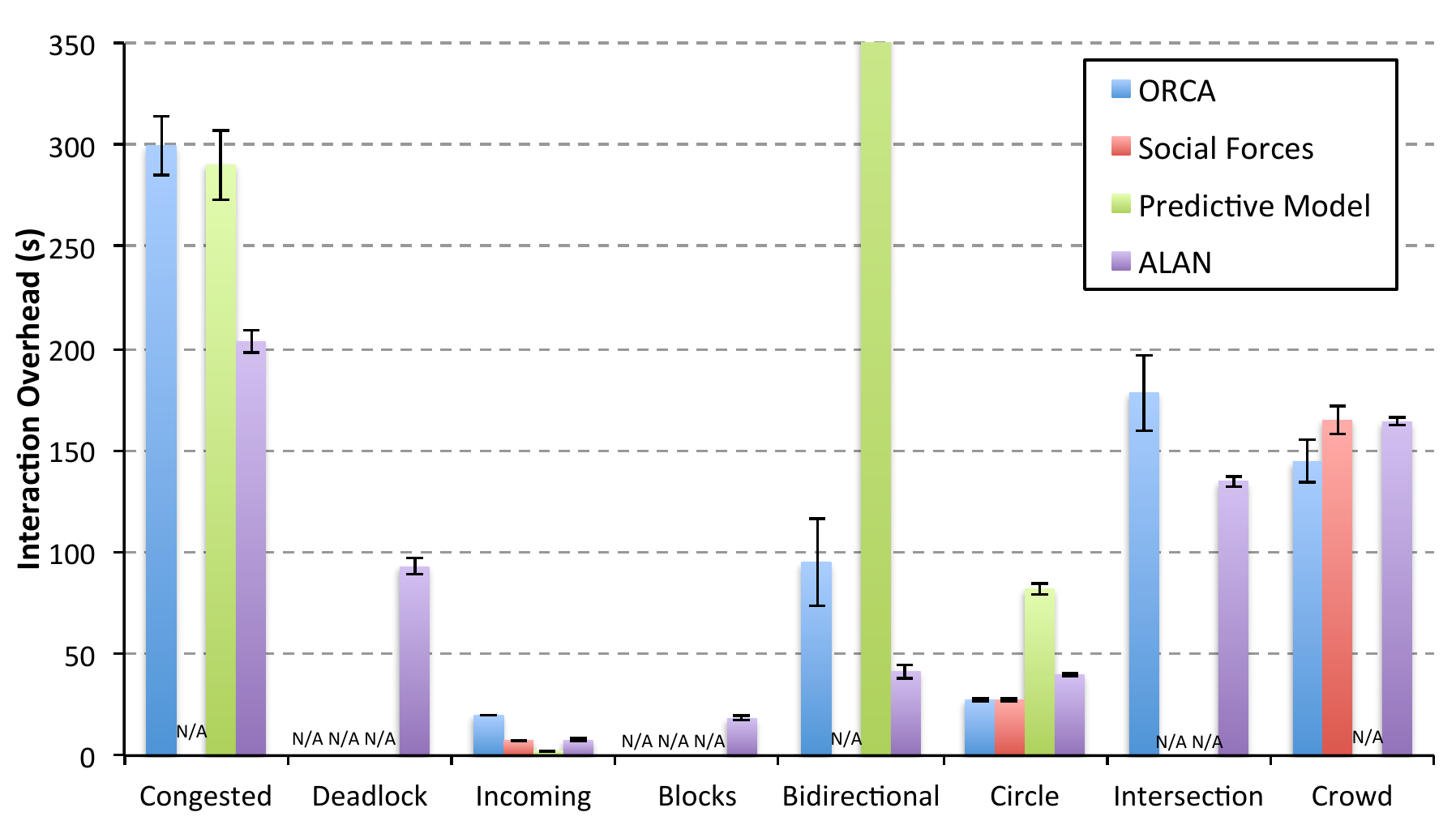}

 \caption{Interaction overhead of ORCA, the Social Forces, the Predictive model, and ALAN in all scenarios. N/A indicates cases where the corresponding method was unable to get agents to their goals.}
 \label{fig:socialforces}
\end{figure}

Results from this comparison can be observed in Figure~\ref{fig:socialforces}. Overall, ALAN outperforms the other approaches in most cases, and gets agents to their goals even when the other three approaches fail to do so. In scenarios with obstacles, ALAN is able to move the agents to their goals, while some (sometimes all) other evaluated approaches cannot. Here, the good performance of ALAN  can be explained by the diversity of motions available and the behavior encouraged by the reward function, which allows agents to find alternative goal paths while avoiding obstacles and, when such paths do not exist (such as in the {\sc Deadlock} scenario) it allows agents to adapt a polite behavior and ``get out of the way" of other agents, backtracking and allowing them to move to their goals.

In obstacle-free scenarios ({\sc Circle} and \textsc{Incoming}), agents have more space to maneuver while moving to their goals. Hence, finding an implicitly coordinated motion is not as critical as in the previous case. In a large scenario, such as \textsc{Circle} the exploratory behavior of ALAN before and after congestion (where agents can move to their goals unconstrained) prevents it from outperforming ORCA and the Social Force models. In a small scenario like \textsc{Incoming}, the overhead of exploration does not affect ALAN as much as in  \textsc{Circle}, allowing it to outperform both ORCA and the Social Force model. However, with the Predictive model, agents in the group make space for the single agent to move directly to its goal, reaching it faster than with ALAN.

From this evaluation, we can observe that ALAN works especially well when agents are highly constrained by both other agents and static obstacles, and its performance advantage is more moderate when agents go through long periods of unconstrained motion.

\subsection{Evaluation of Action Selection Method}
\label{sec:method}

A key component of ALAN is its Softmax inspired action selection method. Here, we validate this design choice by comparing the interaction overhead of different action selection methods, namely, $\epsilon$-greedy~\cite{SB98} (with an $\epsilon$ value of 0.1) and UCB~\cite{auer2002finite}, within the context of ALAN. This evaluation is done using the Sample action set (Fig.~\ref{fig:actions15}).

\begin{figure}[!ht]
 \centering
 \includegraphics[width=0.99\columnwidth]{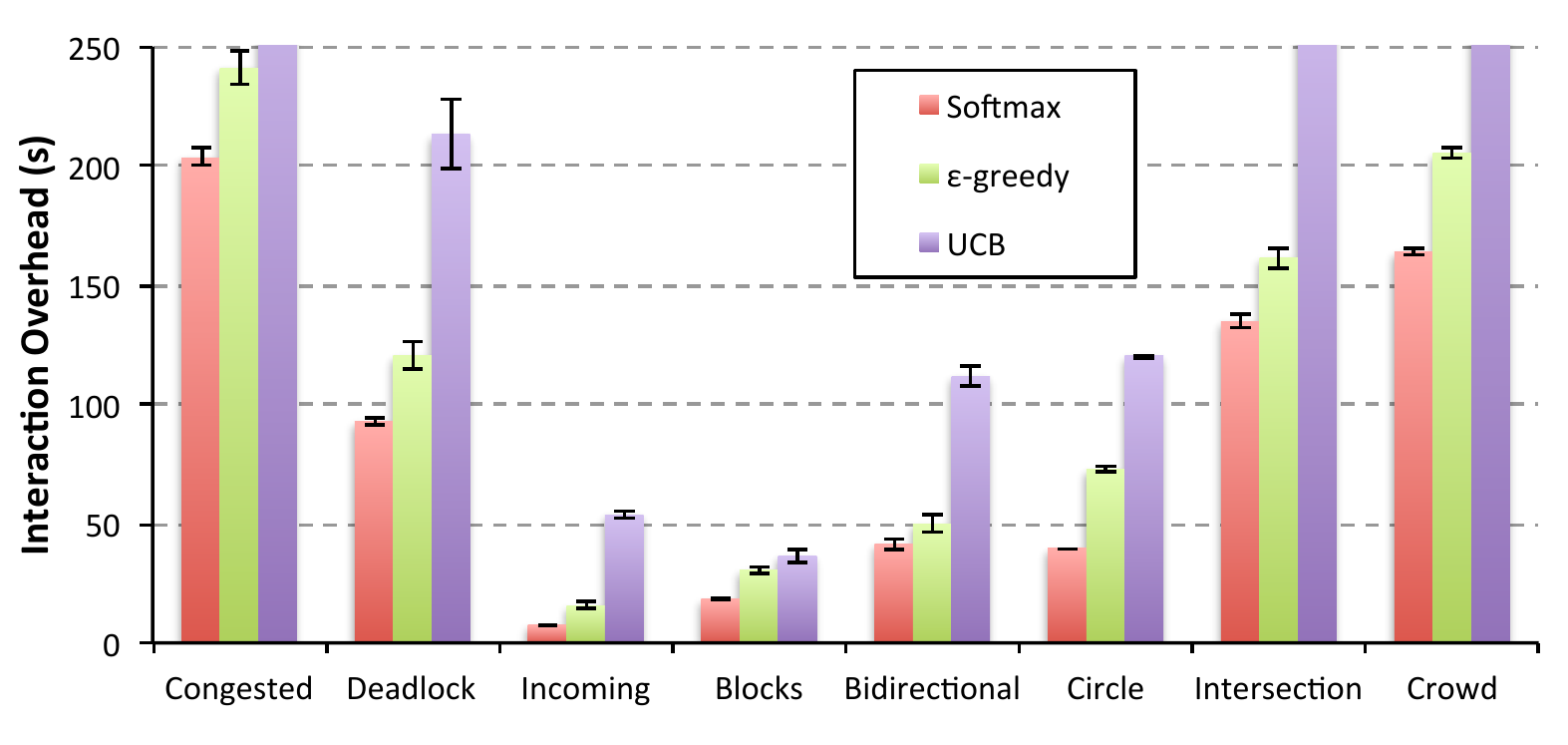}
 \caption{Interaction overhead of ALAN, using different action selection methods (Softmax, $\epsilon$-greedy, and UCB) with the Sample action set (Fig.\ref{fig:actions15}) in all scenarios.}
 \label{fig:actionmethods}
\end{figure}

Results (Fig.~\ref{fig:actionmethods}) indicate that the Softmax action selection helps ALAN achieve the best results. This can be explained by the combination of Softmax's probabilistic nature and its non-uniform randomized exploration. Unlike $\epsilon$-greedy, in Softmax exploration is inversely proportional to action values.
Unlike UCB, the action choice is probabilistic, and it does not depend on the frequency with which each action has been chosen, which is important as that number is not necessarily related to the optimal action.

\subsection{Effect of time window size}
\label{sec:twresult}
\begin{figure}[!ht]
 \centering
 \includegraphics[width=1.0\columnwidth]{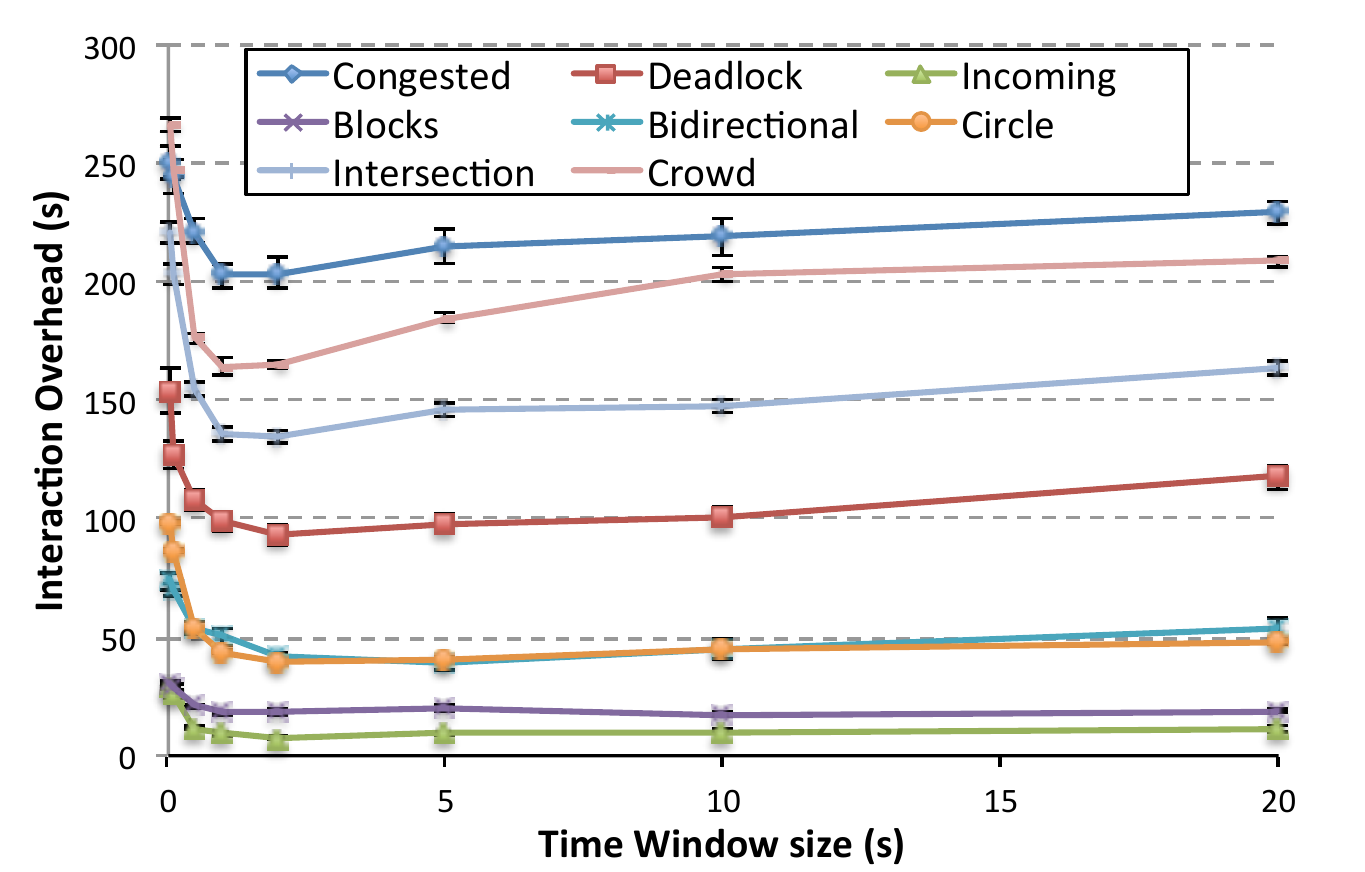}
 \caption{Interaction overhead of ALAN in all scenarios, for different sizes of the time window used for computing the estimated value of each action.}
 \label{fig:tw}
\end{figure}

To answer the question of what should be the time window size in ALAN we show, in Figure~\ref{fig:tw}, a summary of the interaction overhead results obtained by varying the size of the time window (up to 20 secs.).

Figure~\ref{fig:tw} shows that agents perform best when using a time window of approximately 1-5 seconds, which corresponds to approximately $5-25$ action decisions (as a decision is made on average every 0.2 secs). In general, keeping the estimated values for too long or too little time hurts performance. Discarding action estimates too quickly (which turns their value into zero) makes the agent ``forget'' the previously chosen actions. This means that, while exploring, an agent does not have an intuition of which actions can provide a better or worse reward value, as all have the same probability of being chosen. 
On the other hand, keeping action estimates for too long perpetuates possibly outdated values, and reduces the probability of choosing an action that might have recently increased its quality. 
Results show that a time window of 1-5 seconds provides a good balance: it provides agents with some recent information, useful for biasing the exploration towards recently tried ``good" actions and away from ``bad" actions, while also preventing an outdated reward value from introducing noise in the action decision of the agent. 
Unless otherwise noted, we use a time window of 2 seconds throughout all our experiments. 

\subsection{Coordination Factor $\gamma$}
\label{sec:gamma}
\begin{figure}[!ht]
 \centering
 \includegraphics[width=0.95\columnwidth]{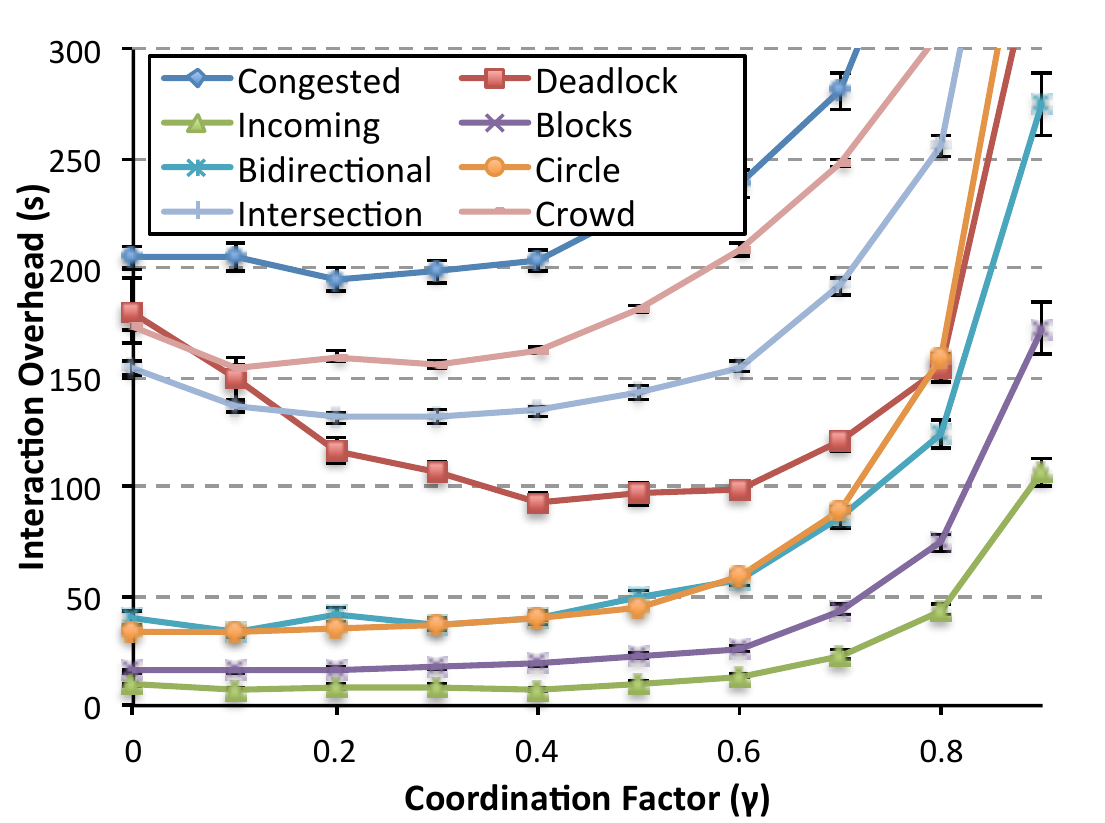}
 \caption{Interaction overhead of ALAN in all eight scenarios, for a range of values of the coordination factor parameter $\gamma$.}
 \label{fig:gamma}
\end{figure}

The coordination factor $\gamma$ controls how goal oriented or polite are the agents in ALAN, based on the reward function (Eq.~\ref{eq:reward}).
Figure~\ref{fig:gamma} shows how the value of $\gamma$ affects the performance of ALAN. We varied the value of $\gamma$ between 0 and 0.9, where $\gamma$=0 means that agents optimize their actions only based on their goal progress, while $\gamma$=0.9 implies that agents optimize their actions based mostly on their politeness, and barely take into account their goal progress. With $\gamma$=1 agents make no progress towards their goal.

A first observation, based on Figure~\ref{fig:gamma}, is that a high weight on the politeness component (a high value of $\gamma$) increases the interaction overhead in all scenarios. This is most noticeable with values of $\gamma>0.6$. Here, the agents are too deferent towards each other, which ends up slowing down their progress. On the other hand, a high weight on the goal oriented component (low values of $\gamma$) seems to only have a significant negative effect on the \textsc{Deadlock} scenario, and a slight negative effect on the \textsc{Intersection}. In the \textsc{Deadlock} scenario, agents on one group are forced to move backwards to exit the narrow corridor. Selfishly maximizing their goal progress prevents agents (of one group) from quickly backtracking and clearing the way for agents in the opposite group. In this case, a balance between goal oriented and polite behavior ($\gamma$ values between 0.3 and 0.6) allows agents to more quickly switch between both types of behavior. In other scenarios, ALAN is robust to a wide variety of $\gamma$ values, minimizing the interaction overhead values when $\gamma<$ 0.5. In these cases, optimizing the action selection based mostly on the goal progress allows agents to find alternative goal paths, using the open space to avoid congestion. 

Overall, giving slightly more weight to the goal oriented component than the politeness component allows agents to alternate between goal oriented and polite behaviors based to their local conditions, showing deference to other agents in order to avoid (or resolve) congestion but also moving to the goal when the path is clear. For these reasons, we used a $\gamma$ value of 0.4 in all ALAN experiments.

\section{Action Space Learning}
\label{sec:mcmc}
The success of the motion scheme in ALAN depends strongly on the action space, i.e. the action set specifying the preferred velocities.  The action selection we have shown is limited to the pre-defined sample set of actions (Fig. \ref{fig:actions15}). However, depending on the environment, different sets of actions might provide motions which improve the navigation. 

We propose an offline learning approach based on a Markov Chain Monte Carlo (MCMC) method \cite{hastings1970monte,metropolis1953equation} with simulated annealing~\cite{kirkpatrick1983optimization} to determine, for a given environment (or set of environments), the set of actions that minimizes the travel time.

Although MCMC is typically used as a sampling method, we use it as an optimization method of sampling with a bias towards regions of better performance. We chose MCMC over other methods because of the nature of the problem, i.e. the effectiveness of any subset of actions depends on the others. First, greedy methods like gradient descent would not be successful given the local minima. Second, the bandit formulation for choosing actions within an action set does not apply because the optimization cannot be decomposed to each action. Third, evolutionary methods only work well when better solutions to subproblems (subsets of the actions) are likely to provide a better solution to the whole problem, which is not our case.

Our method is summarized in Algorithm~\ref{algo:actionspace}. It starts from a set composed of two actions, one action along the goal direction, the other action in a random direction. The MCMC process searches through the action space with biased exploration towards action sets that promote more time-efficient interactions. The explored action set with the highest performance is regarded as the result at the end of the process. Below we describe each step in more detail.

\begin{algorithm}[!ht]
\caption{The MCMC action space learning}
\label{algo:actionspace}
\begin{algorithmic}[2]
\STATE $Act \gets \{GoalDir, RandomDir\},  Act_{opt} \gets Act$
\STATE $F \gets Evaluate(Act),  F_{opt} \gets F$
\STATE $\mathcal{T} \gets \mathcal{T}_{init},  d\mathcal{T} \gets (\mathcal{T}_{final} - \mathcal{T}_{init})/(N - 1)$
\FOR {$i = 1$ to $N$}  %
	\STATE $M \gets SelectModification(Act, i)$
    \STATE $Act' \gets ApplyModification(Act, M)$
    \STATE $F' \gets Evaluate(Act', i)$
    \IF {$ F' < F_{opt}$}
       	\STATE $F_{opt} \gets F',  Act_{opt} \gets Act'$
	\ENDIF
	\IF {$Rand(0, 1) < q(Act, Act')exp((F-F')/\mathcal{T})$}
       	\STATE $F \gets F',  Act \gets Act'$
	\ENDIF
    \STATE $\mathcal{T} \gets \mathcal{T} - d\mathcal{T}$
\ENDFOR
\RETURN $Act_{opt}$
\end{algorithmic}
\end{algorithm}

\textbf{\noindent{Action Set Modification. }} In each iteration, we perform one of the following types of modifications:
\begin{itemize} [leftmargin=2em] \itemsep 0em
\item Modify an action within an interval around its current direction, symmetric on both sides.
\item Remove an action that is not the initial goal-directing one.
\item Add an action within the modification interval of an existing action.
\end{itemize}
\noindent The first type of modification is explored with higher weight (i.e. performed more often),
because we consider the quality of the actions to be more important than the number of actions. Following the simulated-annealing scheme, the modification range decreases over iterations as the simulation moves from global exploration to local refinement. The modification ranges are determined by short learning processes.

\textbf{\noindent{Action Set Evaluation. }}The performance of each new action set is evaluated via ALAN simulation runs. Eq. \ref{eq:ttime} is used to estimate the travel time of the set of agents (here the set of agents is made implicit while the action set is an explicit input to the simulation). We evaluate an action set $Act$ with the function $F$, whose definition is equivalent to the definition of $TTime$ in Eq. \ref{eq:ttime} (Section \ref{sec:man}) but with action set as the explicit argument rather than the set of agents. 

The simulation is repeated multiple times and the average evaluation from all repeated runs is used to evaluate the action set. Following the simulated-annealing scheme, the number of simulation runs increases over iterations, as later local refinement has less uncertainty.

\textbf{\noindent{Action Set Update. }} We use a common version of MCMC, the Metropolis-Hasting Monte Carlo~\cite{hastings1970monte} scheme to reject some of the attempted modifications to efficiently explore better action sets. The probability of keeping a change is related to how it changes the evaluation $F$, which is the key to biasing towards action sets with lower evaluation values. The probability to accept a new action set $Act'$ over a previous action set $Act$ is
\begin{equation}
	min\bigg(1,\ q(Act,Act') \, exp\Big(\frac{F-F'}{\mathcal{T}}\Big)\bigg),
\end{equation}
where $F$ and $F'$ are the evaluation with action set $Act$ and $Act'$ respectively, $q(Act,Act')$ is a factor accounting for the asymmetric likelihood of attempted transitioning between $Act$ and $Act'$, and $\mathcal{T}$ is a parameter within the simulated-annealing scheme. The parameter $\mathcal{T}$ decreases over iterations, making the probability of accepting unfavorable changes decrease, which moves the optimization from global exploration towards local refinement.

After a predefined set of iterations of the MCMC process, the action set $Act$ with the lowest travel time is returned. In our domain, agents have no previous knowledge of the environment, which means that they cannot determine which actions are available beforehand. However, this MCMC approach allows us to do a qualitative analysis of what behaviors are most effective in each type of environment, as we will see in the next Section.

\subsection{Optimized Action Sets}
\label{sec:quali}
To find an optimized set of actions for the scenarios, shown in Figure~\ref{fig:scenarios}, we first learned an optimal action set for each individual scenario.
Then, we used MCMC again to learn an action set that would work well across different scenarios, even ones not considered in the learning process. 

\subsubsection{Action Sets Optimized for Each Scenario}

\begin{figure*}[!ht]
\centering
\hspace*{\fill}%
\subfloat[Congested]{
\includegraphics[width=0.24\textwidth]{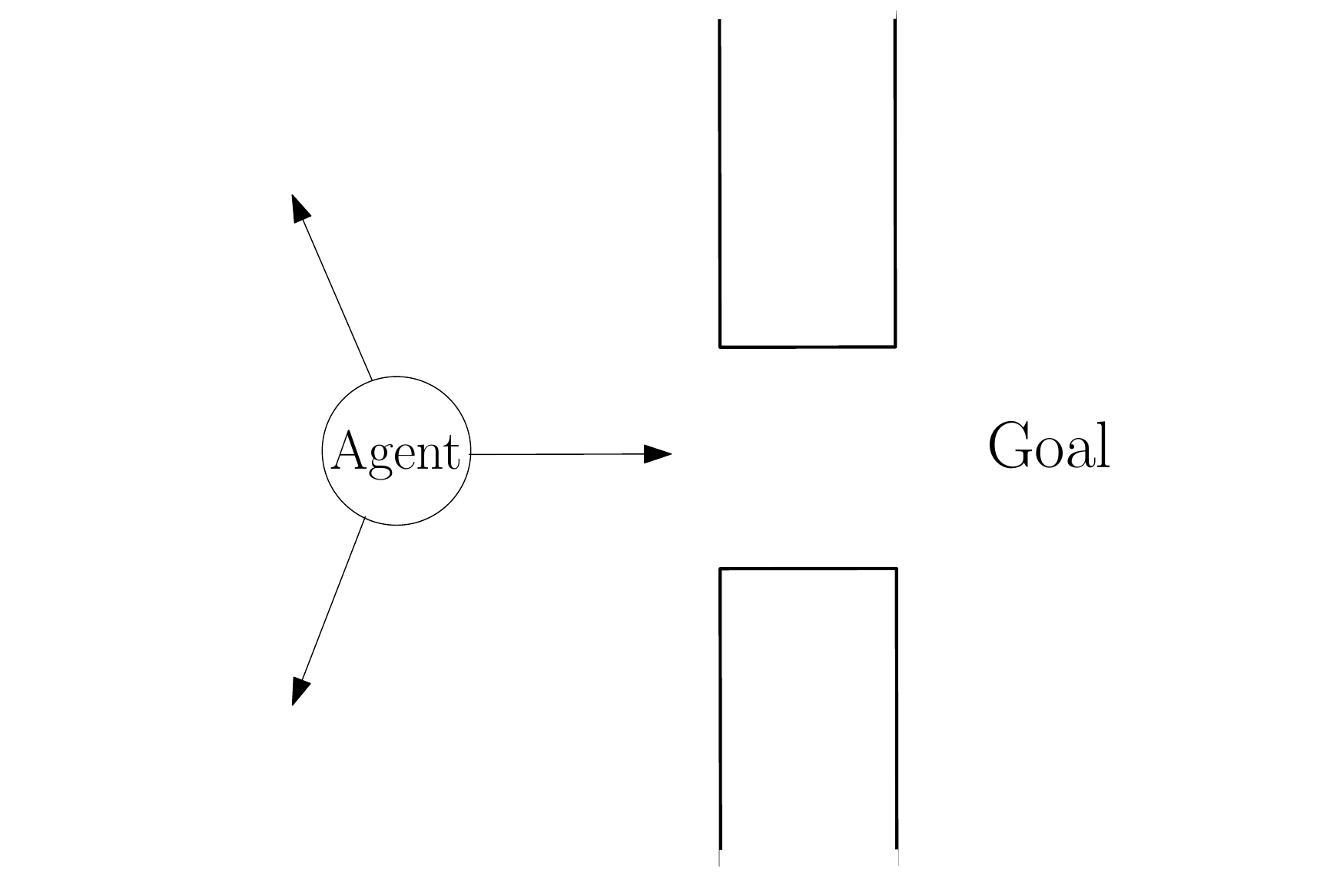}
}\hfill%
\subfloat[Deadlock]{
\includegraphics[width=0.24\textwidth]{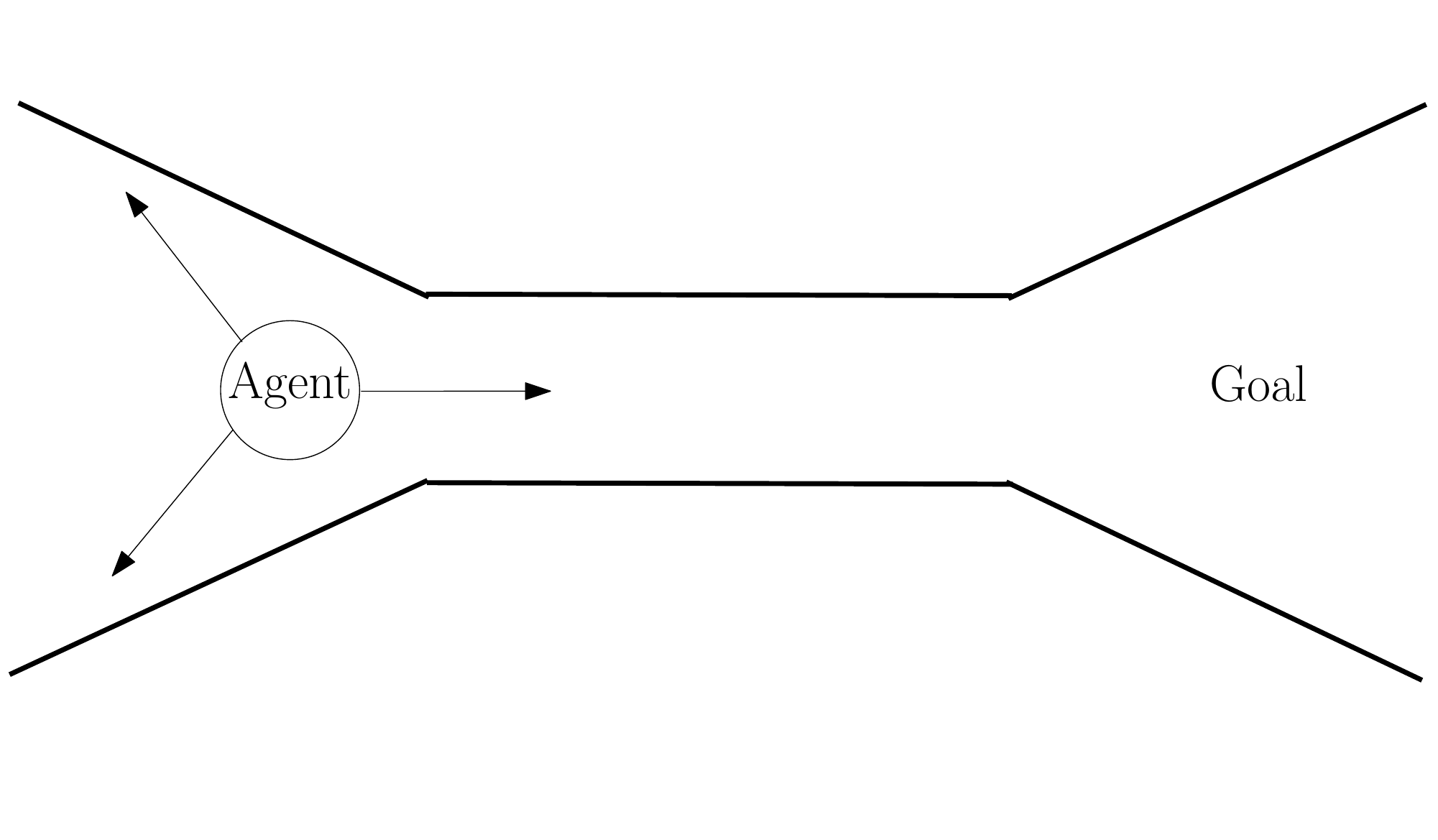}
}\hfill%
\subfloat[Blocks]{
\includegraphics[width=0.24\textwidth]{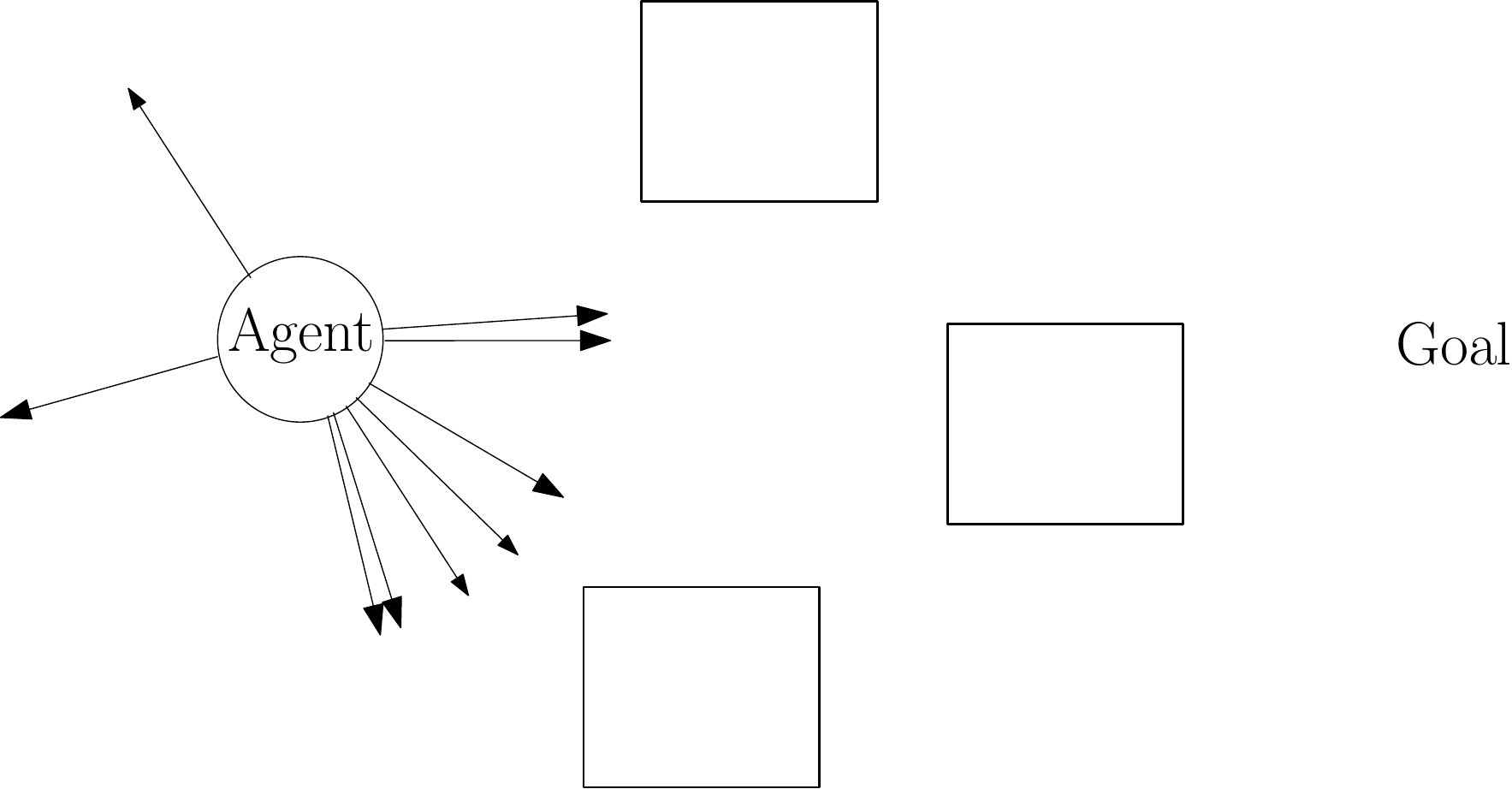}
}\hfill%
\caption{Optimized set of actions found by the MCMC method for the (a) \textsc{Congested}, (b) \textsc{Deadlock} and (c) \textsc{Blocks} scenarios.}
\label{fig:congestedopt}
\end{figure*}

\begin{figure*}[!ht]
\centering
\hspace*{\fill}%
\subfloat[Bidirectional]{
\includegraphics[width=0.22\textwidth]{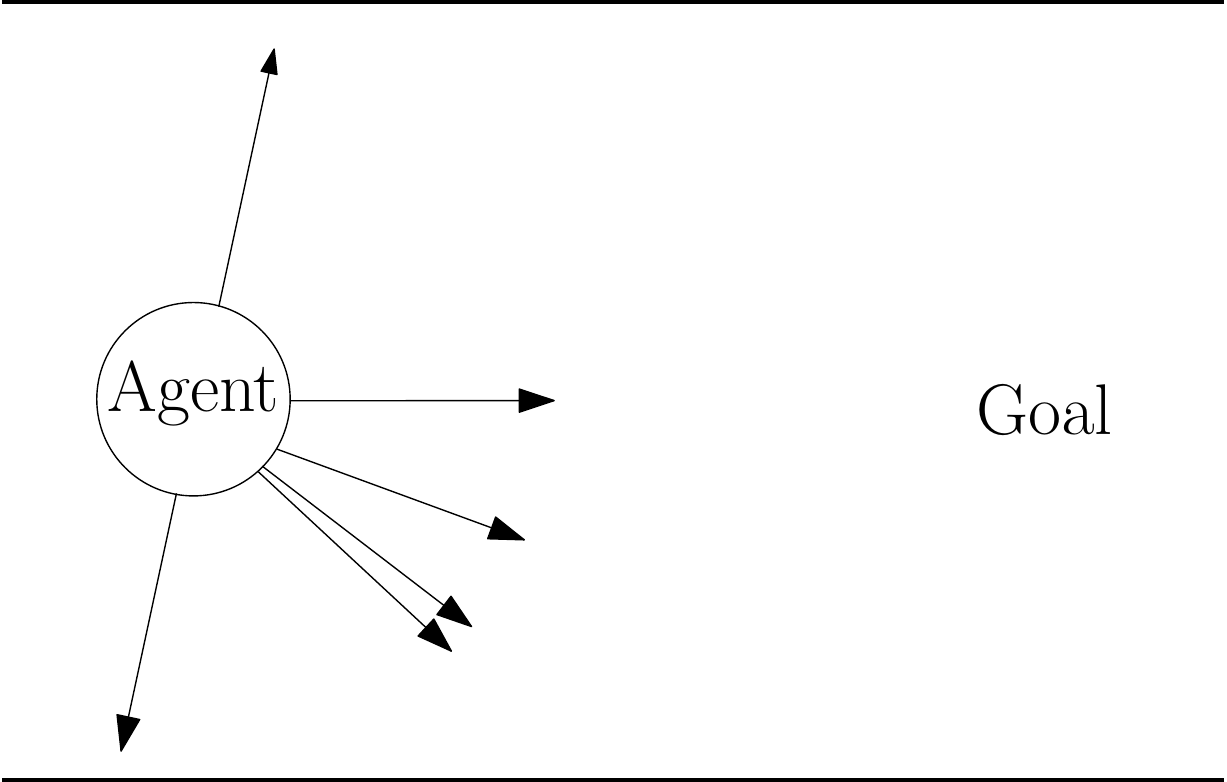}
}\hfill%
\subfloat[Intersection]{
\includegraphics[width=0.2\textwidth]{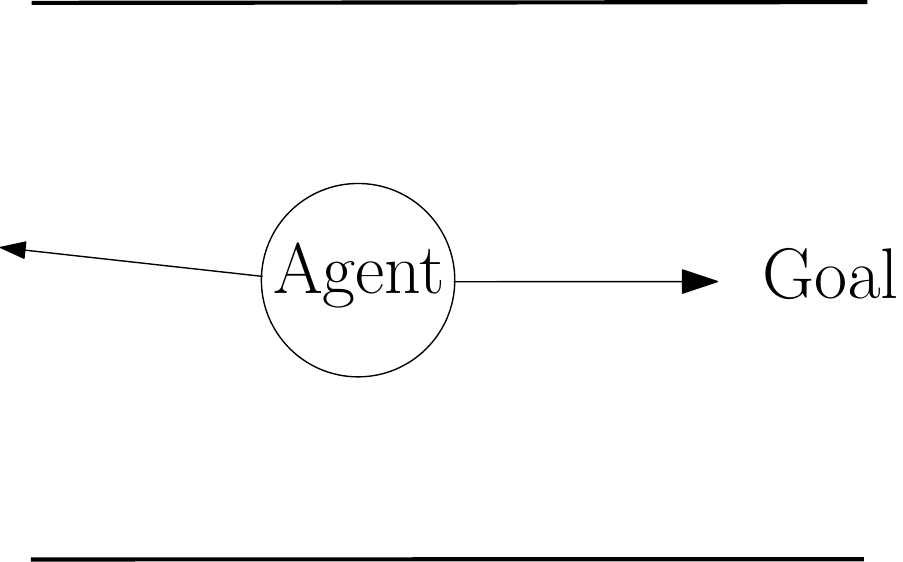}
}\hfill%
\subfloat[Crowd]{
\includegraphics[width=0.2\textwidth]{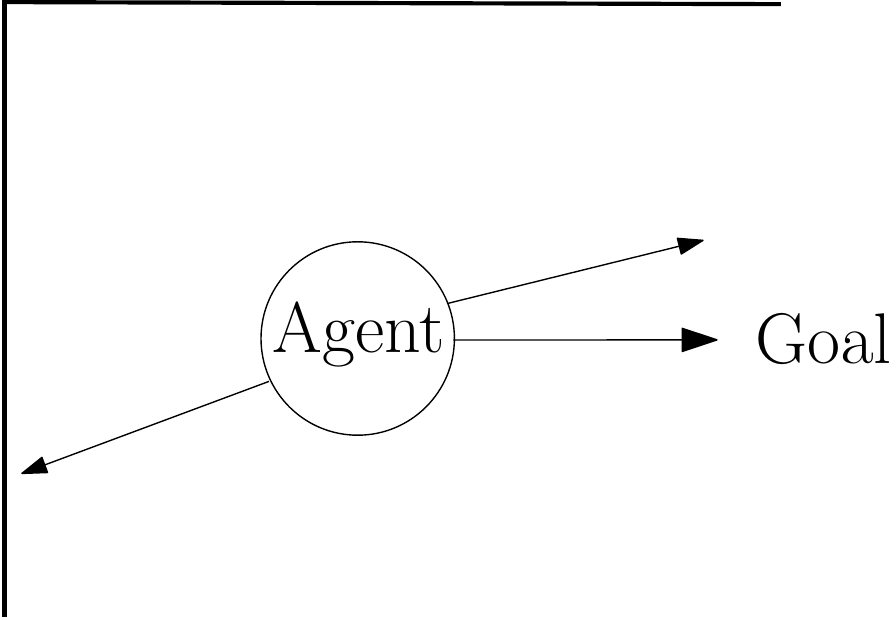}
}\hfill%
\caption{Optimized set of actions found by the MCMC method for the (a) \textsc{Bidirectional}, (b) \textsc{Intersection} and (c) \textsc{Crowd} scenarios.}
\label{fig:congestedopt2}
\end{figure*}

Figures~\ref{fig:congestedopt} and~\ref{fig:congestedopt2} show the set of actions computed by MCMC for the agents in the scenarios: 
\textsc{Congested, Deadlock, Blocks, Bidirectional, Intersection} and \textsc{Crowd}. 

As a general observation, the action set learned for all these scenarios contains at least one action that moves the agent, to some degree, backwards from its goal. 
This backtracking helps in reducing congestion, allowing agents coming from behind, or in the opposite direction, to move to their goals.
In the \textsc{Congested} and \textsc{Deadlock} scenarios, our MCMC approach found  that a set of just 3 actions is enough to minimize the arrival time of the agents. The obstacles that characterize these two environments (in the form of a doorway or a narrow corridor) significantly reduce the flow of agents moving to their goals. In these two scenarios, the backwards actions are almost symmetrical with respect to the goal of the agent. This allows the agent to use the space available outside of the narrow area to spread evenly, left and right, while clearing the path to the goal of the other agents. This behavior avoids congestion on any given side of the narrow area, minimizing the travel time of the agents. 

A set of 3 actions was also found for the \textsc{Crowd} scenario, where the  action that moves the agent sideways helps it to better avoid agents coming in different directions, while the backwards action enables agents to backtrack when there is no space to avoid the congestion that quickly develops in this scenario. 

The action set found in the \textsc{Blocks} scenario is larger (9 versus 3 actions) and highly asymmetrical as compared to the \textsc{Congested} and \textsc{Deadlock} cases. Most actions in this scenario move the agents closer to their goals, unlike the dominant backtracking motions of the previous cases. This indicates that the agents need a more fined grained control to quickly move around the obstacles and reach their goals.

In the \textsc{Bidirectional} scenario, the 6 actions found mostly bias the motion of the agents to their right, similar to the \textsc{Block} scenario. Given the symmetrical nature of this scenario, this bias allows agents to create lanes in each side of the corridor, increasing the efficiency of their own motions and creating space for agents coming in the opposite direction. 

In the \textsc{Intersection} scenario, our MCMC method found that a small set of two actions optimizes the travel time of the agents. The backtracking behavior enabled by the backwards action helps agents resolve congestion in the intersection of the four corridors, which creates space for other agents to move to their goals, reducing overall their travel time.

\begin{figure}[!ht]
\centering
\hspace*{\fill}%
\subfloat[Incoming]{
\includegraphics[width=0.45\columnwidth]{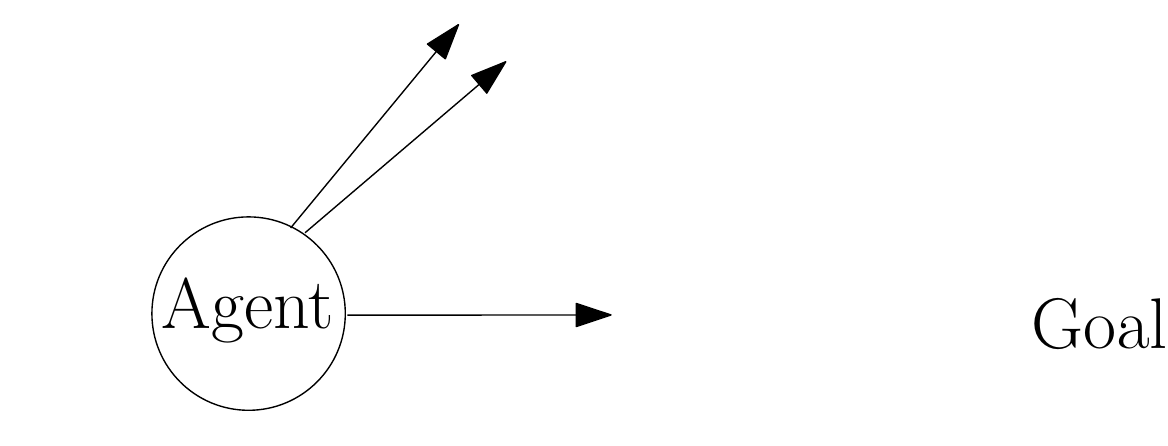}
}\hfill%
\subfloat[Circle]{
\includegraphics[width=0.48\columnwidth]{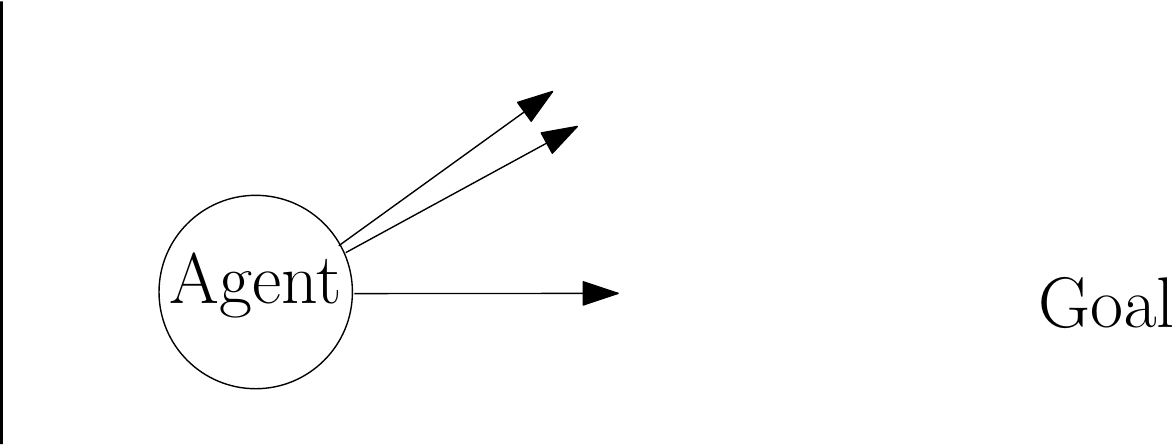}
}
\hspace*{\fill}
\caption{Optimized set of actions (velocities) found by the MCMC method for the (a) \textsc{Incoming} and (b) \textsc{Circle} scenarios.}
\label{fig:incomingopt}
\end{figure}

Figures~\ref{fig:incomingopt}(a) and~\ref{fig:incomingopt}(b) show the optimized set of actions for the  
\textsc{Incoming} and \textsc{Circle} scenarios. A common pattern found by MCMC for these environments is that the actions are heavily biased towards one of the sides of the agents. This bias, along with the absence of obstacles, allows agents to move around other agents using the available space. In the \textsc{Incoming} scenario, the sideway actions help the single agent moving to the right (see Figure~\ref{fig:scenarios}(c)) to avoid getting trapped by the incoming group. 
In the \textsc{Circle} scenario, the optimized actions allow the agents to create a vortex-shaped formation when reaching the center of the environment (see Figure~\ref{fig:scenarios}(f)), which avoids congestion and helps the agents reach their goals faster. Note that, in both scenarios, the two sideways actions are very similar to each other. This gives agents a more fine grained control of their avoidance behavior, minimizing the detour from their goal oriented motion.

\subsubsection{Multi-scenario Optimized Action Set}

To learn a multi-scenario action set, first we trained MCMC on a set of five scenarios, leaving out the \textsc{Bidirectional}, \textsc{Intersection} and \textsc{Crowd} scenarios as test examples. We chose to leave out these  scenarios because without being identical to other scenarios, they share some features with the training set: they have obstacles which constrain the motion of the agents, and also require agents to interact with each other. Then, we evaluated the resulting multi-scenario optimized action set in the entire set of eight scenarios.

\begin{figure}[!ht]
 \centering
 \includegraphics[width=0.5\columnwidth]{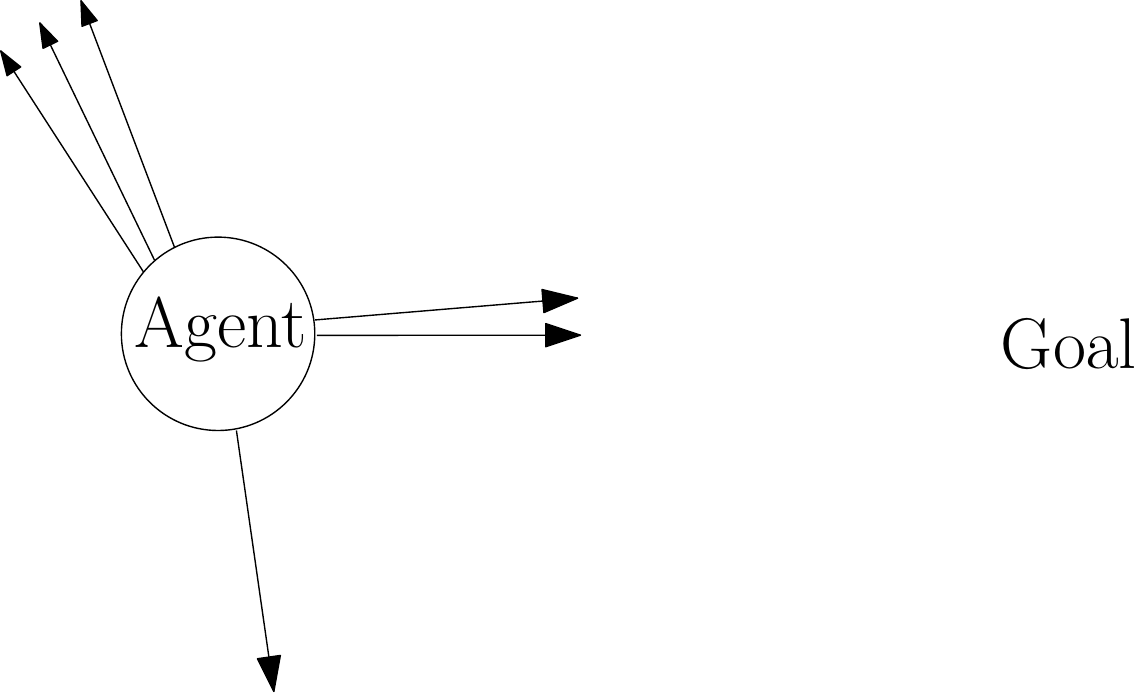}
 \caption{Optimized set of actions (velocities) found by the MCMC method when trained on five of the eight scenarios in Fig.~\ref{fig:scenarios}.}
 \label{fig:globalopt}
\end{figure}

The multi-scenario optimized action set can be seen in Figure~\ref{fig:globalopt}. We can observe two main features of this action set. First, note the asymmetry of the actions, which is helpful in obstacle-free environments to implicitly coordinate the motion of agents and avoid congestion. Second, we can see that half of the actions move the agents backwards from their goals, which is useful in very constrained scenarios. Again, the apparently redundant actions, both backwards as well as towards the goal, give agents better control of their behaviors.

\subsection{Comparison of Performance between Action Sets}

We compared the interaction overhead results of using ALAN with different action sets: the sample set (see Figure~\ref{fig:actions15}), the per-scenario optimized set and the multi-scenario optimized set, computed using our MCMC approach.

\begin{figure}[t]
 \centering
 \includegraphics[width=0.99\columnwidth]{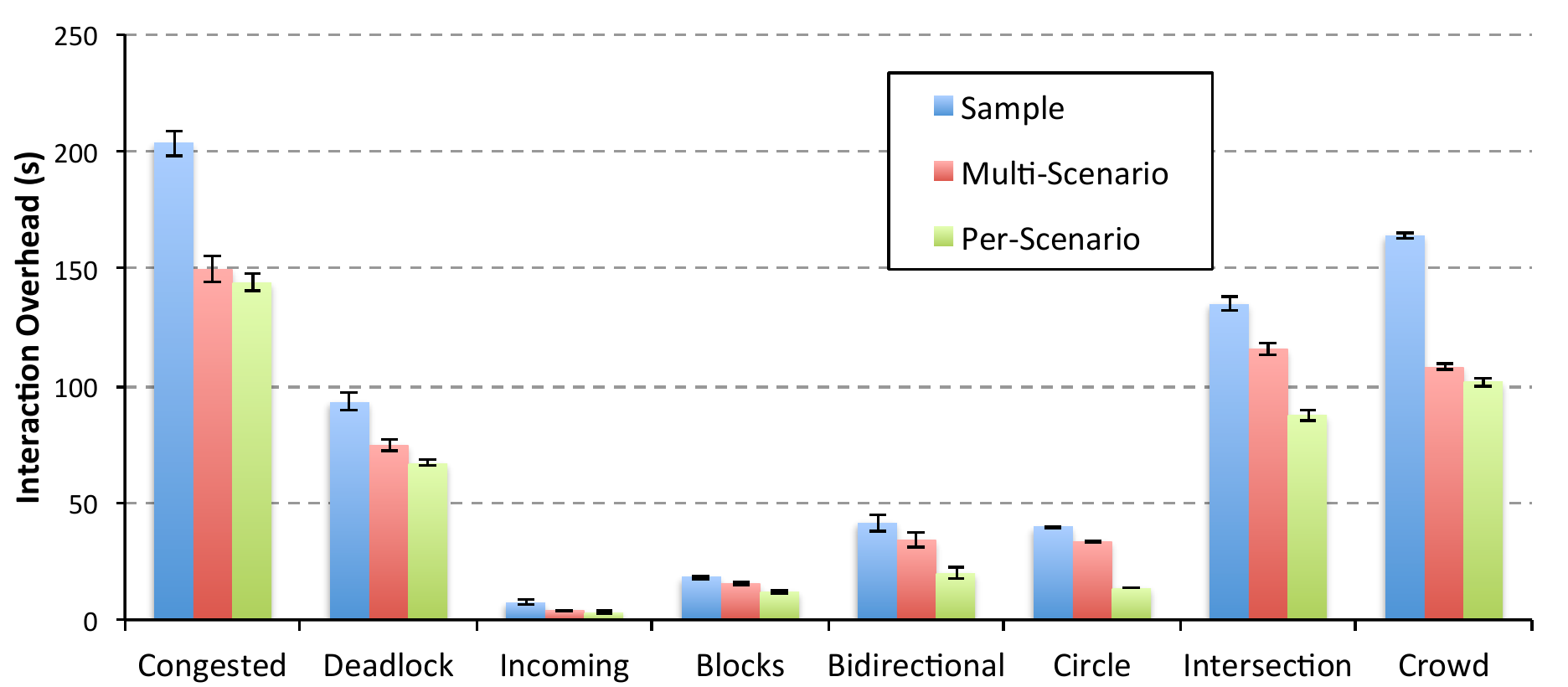}
 \caption{Interaction overhead (s) of ALAN, using the sample action set, the multi-scenario optimized action set, and per scenario optimized action set.}
 \label{fig:opti2}
\end{figure}

Figure~\ref{fig:opti2} shows that our MCMC approach learns optimized action sets that outperform the sample actions. When computed on a per-scenario basis, this optimized action set outperforms the sample action set in all scenarios (in all pairwise t-tests, p $<$ 0.05). When computed using five of the eight evaluated scenarios, it still outperforms the sample set in most scenarios (p $<$ 0.05) while performing similarly in the \textsc{Bidirectional} scenario. Note that while the \textsc{Bidirectional, Intersection} and \textsc{Crowd} scenarios were not included in the training process of the multi-scenario optimized action set, this set still outperforms the Sample action set in the \textsc{Intersection} and \textsc{Crowd}. This indicates that the multi-scenario action set generalizes well to previously unseen environments.

On the other hand, the interaction overhead results of the per-scenario optimized action set are better than the multi-scenario action set, with pairwise differences being statistically significant (p $<$ 0.05) in most scenarios. The only exceptions correspond to the \textsc{Congested} and the \textsc{Incoming} scenarios, where the performance difference is not significant.

We can observe that agents using the multi-scenario optimized action set display behaviors typically attributed to social conventions in human crowds, where pedestrians defer to others to improve the flow and avoid deadlocks. An example of these behaviors can be seen in the \textsc{Deadlock} scenario (backtracking to defer to incoming agents). These behaviors enable agents to reduce their travel time in a wide range of environments without the need for specific (and often unavailable) domain knowledge. Further, the per-scenario optimized action set enables agents to showcase human-like behaviors in the \textsc{Bidirectional} scenario (each group of agents avoids incoming agents moving to their right) and implicitly coordinated motion in the \textsc{Circle} scenario (agents forming a vortex in the middle of the scenario to avoid congestion).

\section{Analysis of ALAN}
\label{sec:sens}

In this section, we analyze different aspects of ALAN, such as its runtime, how its performance scales with respect to the number of agents, as well as its robustness to failure in the actuators of the agents. 
We also compare the performance of ALAN with a strategy where the preferred velocity of each agent is randomized at different time intervals, and show that ALAN outperforms this strategy in all but one scenario. Unless otherwise noted, results labeled with ALAN are obtained with the multi-scenario optimized set of actions (Fig.~\ref{fig:globalopt}).

\subsection{Runtime Complexity}
\label{sec:runtime}
During each simulation cycle, each agent performs two main operations: it first chooses a preferred velocity using its online action-selection algorithm and then maps this velocity to a collision-free one using ORCA. 
In practice, since the number of actions that need to be evaluated is small, selecting a new action has a negligible runtime, while ORCA dominates the overall runtime performance. Consequently, similar to ORCA, ALAN runs in $\mathcal{O} (n)$ time per agent, where $n$ is the number of neighboring agents and obstacles used to compute 
the non-colliding velocity of the agent. 
In time units, ORCA takes approx. $1.5 \times 10^{-5}$ seconds to compute a new collision-free velocity, while ALAN takes approx. $3 \times 10^{-6}$ to select a new preferred velocity. In total, ALAN takes approx. $1.8 \times 10^{-5}$ of processing time for each agent. This corresponds to less than a thousandth of a simulation timestep, which allows us to simulate hundreds of agents in real-time.

\begin{table*}[t]
\setlength{\tabcolsep}{3.5pt}
	\centering
    \begin{tabular}{ | l | r | r | r | r | r | r | r | r |}
    \hline
    Method & \textsc{Congested} & \textsc{Deadlock} & \textsc{Incoming} & \textsc{Block} & \textsc{Bidirectional} & \textsc{Circle} & \textsc{Intersection} & \textsc{Crowd} \\ \hline
    ORCA w/random action every 1s. & 238.2 & 337.5 & 12.7 & 48.2 & \textbf{37.7} & 49.4 & 164 & 151.6   \\ \hline
    ORCA w/random action every 2s. & 290.3 & 553.2 & 10.6 & 106.6 & \textbf{34.4} & 31.8 & 166.2 & 138.5  \\ \hline
    ORCA w/random action every 3s. & 263.4 & 768.9 & 10.9  & 151.9 & \textbf{36.4} & \textbf{28.6} & 129.6 & 137.9 \\ \hline
    ORCA & 299.7 & N/A & 19.8  & N/A & 94.9 & \textbf{27.4} & 178.2 & 144.6\\ \hline
          ALAN & \textbf{149.5} & \textbf{74.4} & \textbf{3.9} & \textbf{15.7} & \textbf{33.9} & 33.4 & \textbf{115.6} & \textbf{107.9} \\
    \hline
    \end{tabular}
    \caption{Comparison of interaction overhead of ORCA, ALAN, and a sample action chosen randomly every few seconds. Bold numbers indicate best results, which may be more than one if there is not a statistically significant difference between them.}
\label{tab:random}
\end{table*}

\subsection{ALAN vs random velocity perturbation}

Table~\ref{tab:random} compares the interaction overhead performance of ALAN, ORCA, and a random action selection, where agents select a random action (from the Sample action set) every 1, 2 or 3 seconds. Results indicate that, in most scenarios, randomizing the selection of the preferred velocity does prevent (or solve) congestion, which results in lower travel times than ORCA in many cases, or even allows agents to reach their goals when ORCA alone cannot. 
Specifically, selecting a random action with some frequency allows agents to reach their goals in the \textsc{Blocks} environment (unlike vanilla ORCA): by randomly moving sideways, agents can escape the local minima behind the obstacles.
We can also observe that in the \textsc{Incoming} and \textsc{Bidirectional} scenarios, the performance is better than just following goal directed motion (ORCA). In the \textsc{Incoming} scenario, specifically, the selection of random actions creates some space between the incoming agents, space that is used by the single agent to (slowly) move to its goal. 

With respect to the performance obtained by ALAN, we can observe that, in general, random perturbations to the preferred velocity of ORCA perform worse than ALAN, as this does not allow agents to adapt to the changes in the local navigation conditions.

\subsection{Scalability}

\begin{figure}[!ht]
\centering
\hspace*{\fill}%
\subfloat[Crowd]{
\includegraphics[width=0.47\columnwidth]{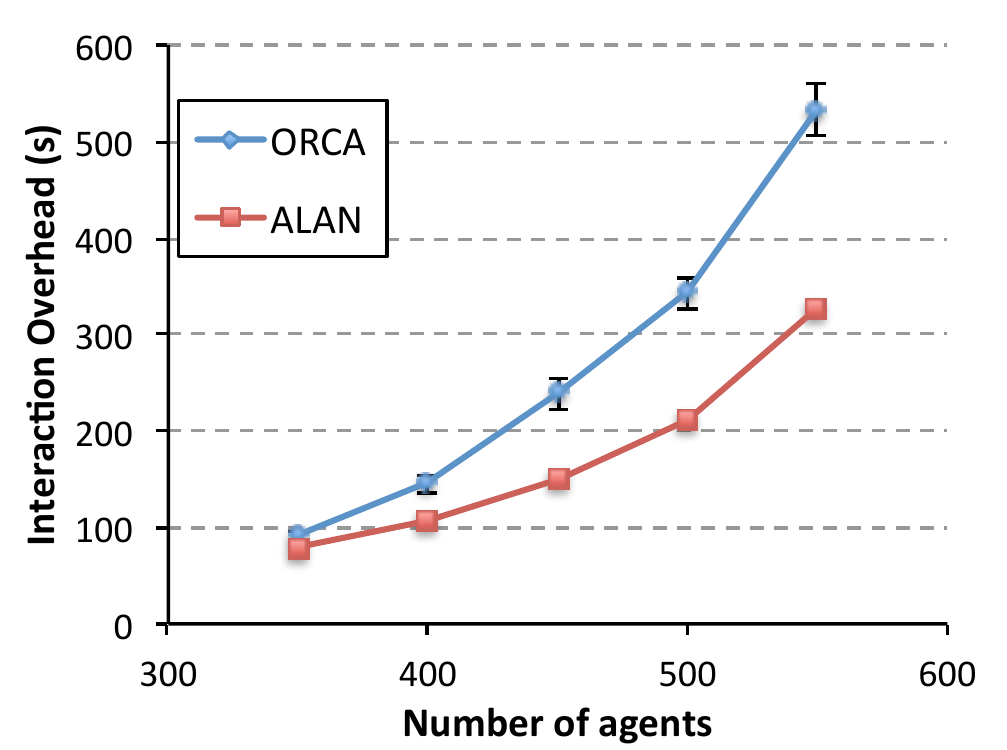}
}\hfill%
\subfloat[Intersection]{
\includegraphics[width=0.47\columnwidth]{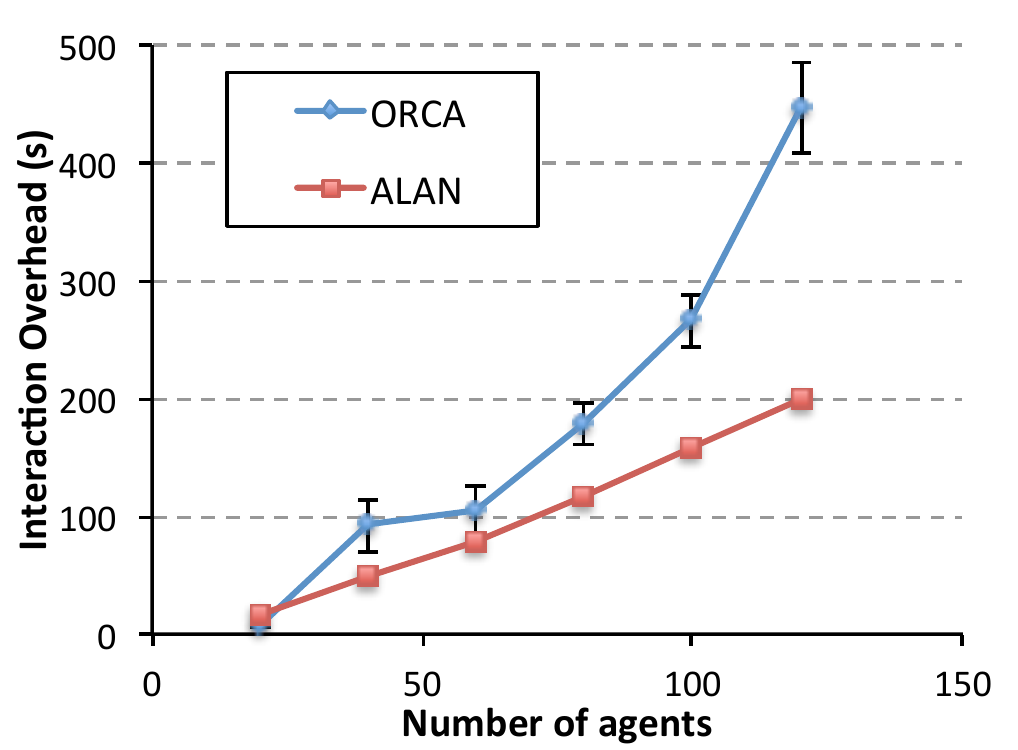}
}
\hspace*{\fill}
\caption{Interaction overhead as a function of the number of agents in the Crowd and Intersection scenarios.}
\label{fig:newscale}
\end{figure}

To analyze how the performance of ALAN scales when there are more agents, we varied the number of agents in the \textsc{Intersection} and \textsc{Crowd} scenarios (Figure \ref{fig:scenarios}), and evaluated the interaction overhead time. Results, shown in Figure \ref{fig:newscale}, indicate that ALAN scales better than ORCA in both scenarios. In the \textsc{Crowd} environment, the performance of ALAN and ORCA is similar with 350 agents. As we increase the number of agents, the difference in interaction overhead is more noticeable. In the \textsc{Intersection} scenario, the difference in performance between ORCA and ALAN is noticeable starting at 40 agents, and increases as the number of agents also increases.

\subsection{Limitations of ALAN}

Although ALAN successfully reduces the travel time of the agents compared to existing navigation approaches, it is not free of limitations. One such limitation relates to the probabilistic nature of its action selection. Specifically, there is no guarantee that agents will always choose the optimal action.  
This can have a negative impact on other agents motions. 
Also, agents evaluate their actions based only on their past observations without considering their long-term consequences. 
This might prevent an agent from reaching its goal, for example, when large obstacles block its goal oriented paths.

\subsubsection{Applicability of ALAN to multi-robot systems}
\label{sec:robot}

To use ALAN in multi-robot systems, some assumptions would need to be changed. Since ALAN uses ORCA for computing collision-free velocities, it makes the same assumptions of holonomic disc-shaped robots as ORCA. ORCA would need to be adapted for computing velocity obstacles for other robot shapes. Currently, we do not assume bounds on the acceleration of the agents and do not consider rotations in the time to take an action. 
Robots with non-holonomic constraints would need to account for rotations and other kinematic constraints, which could be done, for example, using recent extensions to ORCA~\cite{avo,rrvo,mora1}. 
Even without bounds on the acceleration, the motions produced by ALAN look realistic in many of the scenarios, except for ORCA's assumption of 360 degree of sensing range, which is more than double the field of view used by humans. Hence, agents might react to other agents coming from behind to avoid collisions, in a way that might not mimic how humans move.

ORCA assumes that agents have perfect sensing capabilities of the positions and velocities of other agents, which is not necessarily true in decentralized multi robot systems. Fortunately, this problem has been tackled previously by other researchers (for example, \cite{hennes2012multi}, where authors deal with the problem of uncertainty in the localization of agents). Hence, we can use existing solutions to reduce the gap between simulation and real world execution of ALAN. 

\begin{figure}[!ht]
 \centering
 \includegraphics[width=0.9\columnwidth]{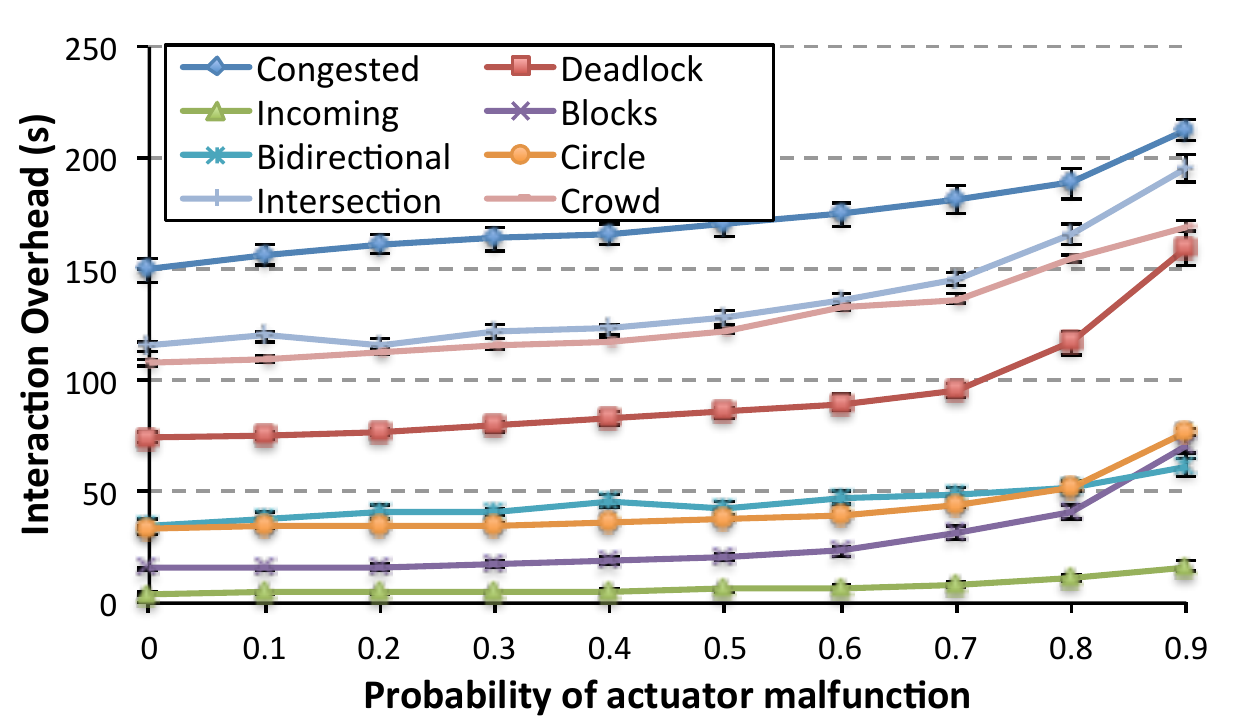}
 \caption{Interaction 
 overhead of ALAN in the eight scenarios shown in Figure \ref{fig:scenarios}, when actions have a probability of not being executed.}
 \label{fig:erroraction}
\end{figure}

\textbf{Imperfect Actuators. } 
We always include a small amount of noise in the computed preferred velocities to avoid symmetry issues. This noise can reflect some level of inaccuracy of the actuators. Since in the real world actuators can fail, we evaluated the performance of ALAN when each action chosen has some probability of not being executed, because the actuators failed. Results, shown in Figure \ref{fig:erroraction}, indicate that ALAN is robust to failure in the actuators. Performance degrades gracefully as the probability of actions not being executed increases. Specifically, the rate at which the interaction overhead values increase depends on the frequency of change of the locally optimal action. In the \textsc{Incoming} scenario, for example, the locally optimal action for the single agent only changes a couple of times (to avoid the group and to resume goal oriented motion), hence the performance degradation is not very noticeable until the probability of actuator failure is over 70\%. On the other hand, in the \textsc{Congested} scenario the performance degradation is visible at around 20\% of probability of actuator failure. Overall, this result indicates that ALAN still performs well under these conditions.

\section{Conclusions and Future Work}
\label{sec:conc}

In this paper, we addressed the problem of computing time-efficient motions in multi-agent 
navigation tasks, where there is no communication or prior coordination between the agents. 
We proposed ALAN, an adaptive learning approach for multi-agent navigation. We 
formulated the multi-agent navigation problem as an action selection problem in a multi-armed bandit setting, 
and proposed an action selection algorithm to reduce the travel time of the agents. 
 
In  ALAN, the agents use recent observations to decide which action to execute. 
ALAN uses principles of the Softmax action selection strategy and a limited time window of rewards to dynamically 
adapt the motion of the agents to their local conditions. 
We also introduced an offline Markov Chain Monte Carlo method that allows agents to learn an optimized action space in 
each individual environment, and in a larger set of scenarios. This enables agents to reach their goals faster than using a predefined set of actions.

Experimental results in a 
variety of scenarios and with different numbers of agents show that, in general, agents using ALAN 
make more time-efficient motions than using ORCA, the Social Forces model, and a Predictive model for pedestrian navigation. ALAN's low computational complexity and completely distributed nature make it an ideal choice for  multi-robot systems that have to operate in real-time, often with limited processing resources.
 
There are many avenues for future research. We plan to investigate the applicability of ALAN 
to heterogeneous environments, for example, by letting ALAN agents 
learn the types of the other agents present in the environment and their intended goals.  
This would allow an agent to more accurately account for the behavior of nearby agents  
during action selection. Finally, we would also like to port our approach to real robots and test it in real-world
environments, such as for search and rescue operations or evacuation planning.

\newpage

\end{document}